\documentclass[superscriptaddress,reprint,amsmath,amssymb,floatfix]{revtex4-2}
\usepackage{graphicx}  
\usepackage[T1]{fontenc}
\usepackage{siunitx}        %For unit representation
\usepackage{wrapfig}
\usepackage{float}
\usepackage{placeins}
\usepackage{xcolor}
\usepackage{makecell}       %For tables

	\setcellgapes{4pt}

\usepackage{mathtools}
\usepackage{enumitem}
\usepackage{lmodern}
\usepackage{microtype}
\usepackage{nicefrac}   % for inline fraction 
\usepackage{xspace}         %to include conditional spaces in commands
\usepackage{rotating}
\usepackage{etoolbox}

\newcommand{\affila}{Laboratory for Solid State Physics, ETH Zurich, 8093 Zurich, Switzerland}
\newcommand{\affilb}{European XFEL, Holzkoppel 4, 22869 Schenefeld, Germany}
\newcommand{\affilc}{Institute of Physics, University of Kassel, 34132 Kassel, Germany}
\newcommand{\affild}{Wilhelm Ostwald Institute for Physical and Theoretical Chemistry, Leipzig University, 04103 Leipzig, Germany}
\newcommand{\affile}{Quantum Center of Excellence for Diamond and Emergent Materials and Department of Physics, Indian Institute of Technology Madras, Chennai 600036, India}
\newcommand{\affilf}{Institute of Physics, University of Rostock, 18059) Rostock, Germany}
\newcommand{\affilg}{PNSensor GmbH, 81739 Munich, Germany}
\newcommand{\affilh}{Physics Department, University of Connecticut, Storrs, Connecticut 06269-3046, United States}
\newcommand{\affili}{Institute for Optics and Atomic Physics, Technical University Berlin, 10623 Berlin, Germany}
\newcommand{\affilj}{Indian Institute of Science Education and Research Pune, Pune 411008, India}
\newcommand{\affilk}{Max Planck Institute for Nuclear Physics, 69117 Heidelberg, Germany}
\newcommand{\affill}{Institute of Physics, University of Freiburg, 79104 Freiburg, Germany}
\newcommand{\affilm}{Kimika Fakultatea, Euskal Herriko Unibertsitatea (UPV/EHU) and Donostia International Physics Center (DIPC), P.K. 1072, 20080 Donostia, Spain}
\newcommand{\affiln}{IKERBASQUE, Basque Foundation for Science, 48011 Bilbao, Spain}
\newcommand{\affilo}{Department of Chemistry, Aarhus University, 8000 Aarhus C, Denmark}
\newcommand{\affilp}{Center for Interdisciplinary Nanostructure Science and Technology (CINSaT), University of Kassel, 34132 Kassel, Germany\\
Corresponding authors: Linos Hecht (\texttt{lhecht@phys.ethz.ch}) \\    Yevheniy Ovcharenko (\texttt{yevheniy.ovcharenko@xfel.eu})\\ Daniela Rupp (\texttt{ruppda@phys.ethz.ch})\\ Marcel Mudrich (\texttt{mudrich@uni-kassel.de})}

\usepackage{hyperref}						%for shortcuts and links
\hypersetup{colorlinks=false,   % Keep text the same color as the body
            pdfborder={0 0 0},  % Remove the colored boxes around links
            breaklinks=true}

\begin{document}

\title{Model-free pattern separation of two-color ultrafast X-ray diffraction}

\author{Linos \surname{Hecht}}\affiliation{\affila}
\author{Yevheniy \surname{Ovcharenko}}\affiliation{\affilb}

\author{Asbjørn Ø. \surname{Lægdsmand}}\affiliation{\affilc}
\author{Björn \surname{Bastian}}\affiliation{\affild}
\author{Thomas M. \surname{Baumann}}\affiliation{\affilb}
\author{Alessandro \surname{Colombo}}\affiliation{\affila}
\author{Subhendu \surname{De}}\affiliation{\affile}
\author{Alberto \surname{De Fanis}}\affiliation{\affilb}
\author{Simon \surname{Dold}}\affiliation{\affilb}
\author{Thomas \surname{Fennel}}\affiliation{\affilf}
\author{Robert \surname{Hartmann}}\affiliation{\affilg}
\author{Katharina \surname{Kolatzki}}\affiliation{\affila}
\author{Sivarama \surname{Krishnan}}\affiliation{\affile}
\author{Björn \surname{Kruse}}\affiliation{\affilf}
\author{Aaron C. \surname{Laforge}}\affiliation{\affilh}
\author{Bruno \surname{Langbehn}}\affiliation{\affili}
\author{Suddhasattwa \surname{Mandal}}\affiliation{\affilj}
\author{Tommaso \surname{Mazza}}\affiliation{\affilb}
\author{Cristian \surname{Medina}}\affiliation{\affilk}
\author{Christian \surname{Peltz}}\affiliation{\affilf}
\author{Thomas \surname{Pfeifer}}\affiliation{\affilk}
\author{Björn \surname{Senfftleben}}\affiliation{\affila}\affiliation{\affilb}
\author{Keshav \surname{Sishodia}}\affiliation{\affile}
\author{Frank \surname{Stienkemeier}}\affiliation{\affill}
\author{Rico Mayro P. \surname{Tanyag}}\affiliation{\affilo}
\author{Paul \surname{T\"ummler}}\affiliation{\affilf}
\author{Sergey \surname{Usenko}}\affiliation{\affilb}
\author{Andreas \surname{Heidenreich}}\affiliation{\affilm}\affiliation{\affiln}
\author{Michael \surname{Meyer}}\affiliation{\affilb}
\author{Daniela \surname{Rupp}}\affiliation{\affila}
\author{Marcel \surname{Mudrich}}\affiliation{\affilc}\affiliation{\affilp}
\makeatletter
\def\Dated@name{}%
\makeatother

\begin{abstract}
Two-color X-ray imaging with Free Electron Laser pulses offers a powerful approach for probing ultrafast structural dynamics in nanoscale systems, combining (near-)atomic spatial resolution with femtosecond temporal precision. The first X-ray pulse captures the object's initial state, while the second, time-delayed pulse records its subsequent evolution. A key challenge lies in disentangling the two patterns simultaneously recorded by the same detector. We demonstrate the realization of this approach on structurally varying nanoscale particles using two X-ray pulses of different photon energies, 1 and 1.2~keV. Sub-micrometer helium nanodroplets generated in vacuum are irradiated by the two X-ray pulses separated in time by up to 750 femtoseconds.
Taking advantage of the high photon-energy resolution of the imaging detector, we separate the overlapping diffraction signals by analyzing individual pixel counts and applying pattern recognition. The helium nanodroplets' spherical shape allows us to cross-validate this approach by fitting the radial scattering profiles with Mie solutions for abichromatic field. The excellent agreement between the two methods, particularly in the sparsely illuminated outer regions of the diffraction patterns where high-resolution structural information is encoded, highlights the quality of this approach and its potential for future advanced X-ray movie techniques.
\end{abstract}

\maketitle

\thispagestyle{empty}

\section*{Introduction}
Coherent diffraction imaging (CDI) using intense X-ray pulses produced by Free Electron Lasers (FELs) is a powerful technique to determine the size and structure of individual unsupported nanoparticles~\cite{chapmanFemtosecondDiffractiveImaging2006, bielecki2020, sun2022}. Fragile nanosystems such as single living cells~\cite{kimura2014imaging}, viruses~\cite{seibertSingleMimivirusParticles2011} and proteins~\cite{ekeberg2024}, aerosol nanoparticles~\cite{boganSingleParticleXray2008, lohFractalMorphologyImaging2012, colombo2025}, and superfluid helium nanodroplets (HNDs)~\cite{gomezSizesLargeHe2011,gomezShapesVorticitiesSuperfluid2014,tanyag2015communication,jonesCoupledMotionXe2016,tanyagCHAPTERExperimentsLarge2017,o2020angular,feinberg2021aggregation,feinbergXrayDiffractiveImaging2022, ulmer2023} have been successfully investigated with this snapshot-imaging technique.

The femtosecond duration of XFEL pulses further allows for capturing and tracking ultrafast structural dynamics of single nanoparticles via pump–probe excitation schemes. Various types of photoinduced dynamics of nanoparticles have been studied, including the expansion and fragmentation of rare-gas nanoclusters~\cite{gorkhoverFemtosecondAN2016,fluckigerTimeresolvedXrayImaging2016,langbehnDiffractionImagingLight2022}, anisotropic HND surface broadening~\cite{bacellarAnisotropicSurfaceBroadening2022}, melting, bubble-like expansion, and explosion of superheated metal nanoparticles~\cite{Shin2023, doldMeltingBubblelikeExpansion2025}. Matching the observed structural changes with advanced large-scale simulations provides unique insight into non-equilibrium dynamics and allows for the extraction of previously inaccessible properties of highly excited matter \cite{Peltz2022, Shin2023, doldMeltingBubblelikeExpansion2025}.

However, only when the initial state of the nanotarget is precisely known, an image of the evolved nanostructure alone is sufficient to determine the underlying dynamics with confidence. Most types of nanoparticles exhibit some degree of variability in size, shape, and orientation, which obstructs matching simulation and experiment and thus severely limits the reliability of conclusions. To obtain a genuine X-ray movie, two distinct images of the nanotarget's initial and final states must be recorded. 
We identify two particularly advantageous scenarios for this approach. The first involves visualizing the dynamics induced by an intense X-ray pulse in a nanoparticle. This serves as an ideal testbed for refining our fundamental understanding of ultrafast radiation damage \cite{Ziaja2012, ho2020role}. In this setup, the initial X-ray pulse acts as both the pump to induce the dynamics and the probe for the initial state, while a delayed second X-ray pulse probes the final state.
The second scenario focuses on creating movies of laser-driven dynamics, such as plasma formation or shock waves, which are highly relevant for ultrafast material science, EUV lithography, and fusion. This application would require a probe–pump–probe scheme, where the first X-ray pulse ideally has a negligible influence on the initial dynamics.

These visions have motivated substantial efforts to develop and implement two-color X-ray imaging schemes. Advances in FEL technology now enable the generation of intense, independently tunable two-color X-ray pulses \cite{lutman2013, allaria2013a, hara2013, marinelli2015, serkez2020opportunities, prat2022}. Since the temporal delay between the two consecutive pulses is extremely short, fast switching of detectors to record the two images separately is impossible. Therefore, various methods to disentangle the two images of the same nanoparticle have been explored, including spectral filtering at the detector~\cite{ferguson2016,hecht2018b} and spatial separation onto two different detectors~\cite{sauppe2024double}. These approaches on the experimental side, however, all suffer substantial photon losses during the separation process, limiting their efficiency.

\begin{figure*}[!ht]
    \centering
    \includegraphics[width=\textwidth]{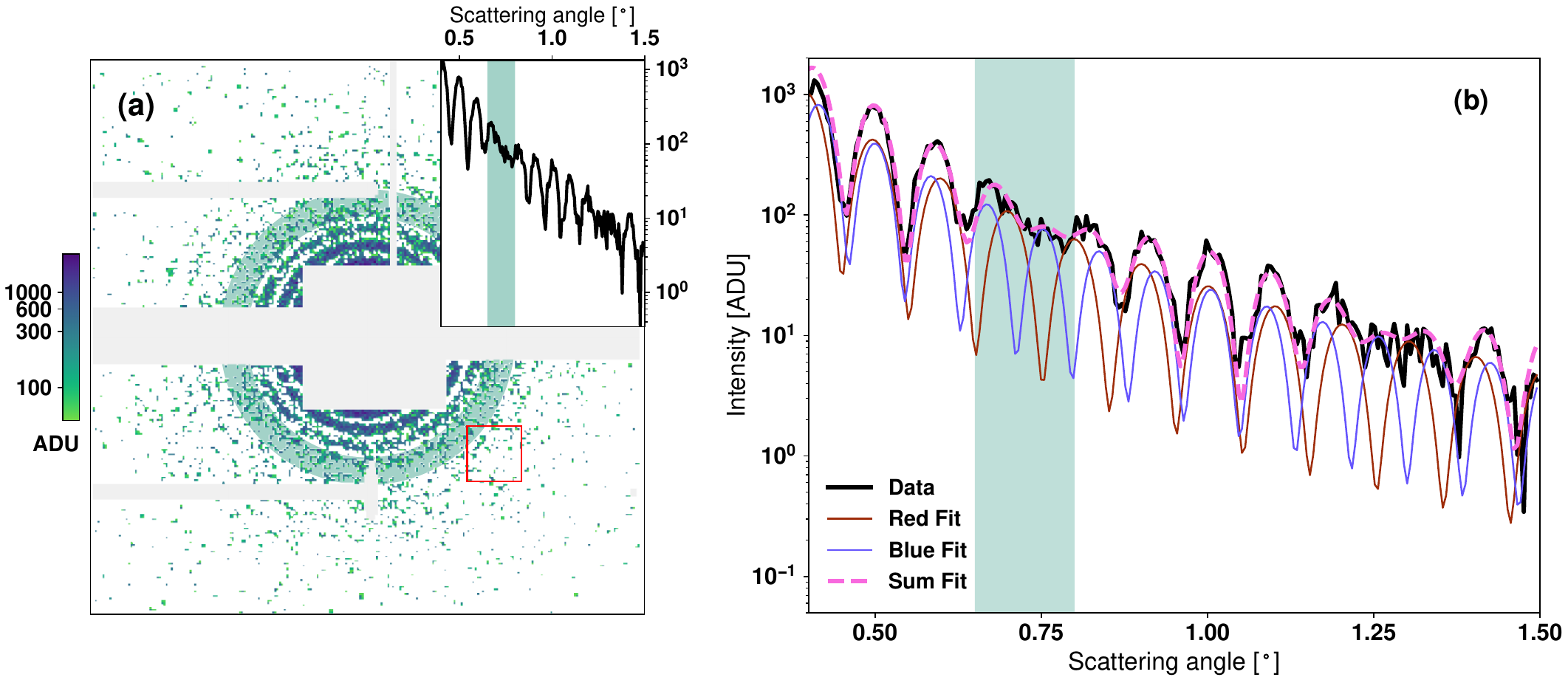}
    \hfill\null
    \caption{(a) Typical coherent diffraction pattern of a pure helium nanodroplet of radius $R \approx354$~nm irradiated by two X-ray pulses with mean photon energies of $992$~eV ``red'' pulse) and $1192$~eV (``blue'' pulse) at nominal FEL intensities of $4.2\times10^{15}\:$~W/cm$^2$ and $1.9\times10^{16}\:$~W/cm$^2$, respectively. The light gray horizontal stripe and the square in the center of the image represent open areas of the detector necessary to prevent damage from the transmitted beam. Additionally, smaller light gray areas are damaged pixels excluded from the analysis. The signal intensity in each pixel is expressed in analog-to-digital units (ADU). The radial pixel intensity profile (black line) is presented as an inset in the first quadrant of Fig. 1(a). It is derived by angular averaging around the center of the diffraction pattern.
    The turquoise-shaded area is marked in the radial profile and highlights the intensity beating due to the components' different frequencies. A close-up view of the red square is presented in the Methods section.
    The magnified view of the radial profile is shown in (b). The result of fitting the experimental profile with Mie solutions for a bichromatic field is shown as a pink dashed line. The solid red and blue lines represent the components of the red and blue X-ray pulses, as determined by the fits.
    \label{fig:full}}
\end{figure*}

Here, we introduce a computational image separation method based on post-processing. It leverages the ability of the European XFEL (EuXFEL) to generate two co-propagating X-ray pulses with distinct photon energies~\cite{serkez2020opportunities}. The experimental realization relies on a specialized pn-junction charge-coupled device (pnCCD) imaging detector at the Small Quantum Systems instrument. Notably, this detector provides a spectral resolution of about 10\% of the photon energy alongside low readout noise~\cite{struder2010, kuster20211}. Individual nanoparticles are illuminated with two X-ray pulses at different photon energies (1 and 1.2 keV), separated in time by up to 750 femtoseconds.

As an ideal model target system, we use sub-micrometer HNDs, produced by expanding liquid helium into vacuum through a small nozzle~\cite{ulmer2023}. HNDs exhibit a broad size distribution and some variety of shapes, but predominantly have a perfect spherical shape~\cite{gomezShapesVorticitiesSuperfluid2014, langbehnThreeDimensionalShapesSpinning2018}. 
Moreover, their absorption cross-section in the X-ray regime is very small. This allows us to treat them as nearly static targets that remain structurally unchanged until the arrival of the second X-ray pulse, enabling detailed benchmarking of our image separation method.

The two superimposed diffraction patterns recorded within the same frame by the pnCCD detector are separated by analyzing individual pixel counts. Due to the spherical shape of the HNDs we can cross-validate the pixel-based color-separation approach by fitting radial profiles of the diffraction patterns with two-color Mie simulations, showing excellent agreement between the two methods.

Our results, supported by molecular dynamics simulations, confirm that indeed, within the spatial resolution of the experiment, no structural damage occurs in the HNDs up to the maximum delay of 750~fs.
Systematic scattering simulations are used to test the limitations of the pixel-wise color separation method. The validity of the assignment is excellent in image regions where the average area density of photon hits on the detector remains below 0.1 detected photons per pixel. For higher photon densities, the spillover of charges across pixel borders and detector noise imposes limitations.

The demonstrated high reliability of our color separation method in sparsely illuminated regions is a particular advantage for two-color single-particle CDI, as these large-momentum-transfer regions encode the crucial high-resolution information. Our pixel-based approach is ideally suited for low-signal applications, such as 3D protein imaging from faint single-shot patterns. For other diffraction-based methods, such as Bragg CDI, the method-specific spatial separation of signals from different X-ray energies will even allow a straightforward extension of our pixel-based separation method to much higher photon densities. In these methods, different X-ray energies inherently diffract into distinctly different regions of the detector. This natural spatial separation prevents multi-color photon pile-up within individual pixels that typically complicates analysis. Consequently, our pixel-based separation could reliably process these data even at much higher overall photon densities. In this regard, our approach extends well beyond two-color single-particle CDI, offering a key building block for future advanced X-ray movie techniques based on two-color XFEL pulses.

\section*{Results}

\begin{figure}[tb]
\centering
\includegraphics[width=\columnwidth]{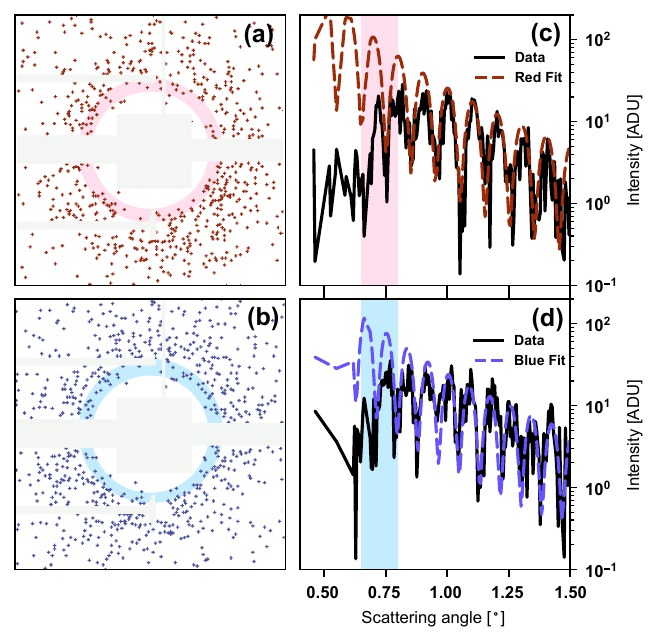}
\caption{(a) Contribution of the red pulse to the image in Fig.~\ref{fig:full}, obtained via pixel-based separation. (b) Corresponding contribution of the blue pulse. Assigned pixels are marked with small crosses; the marker color indicates the assigned pulse color and does not encode intensity. This method is only effective in the outer regions of the image (scattering angle $\theta\gtrsim 0.8\,^\circ$), where lit pixels are sparse, and the likelihood of clusters of lit pixels overlapping is low. (c) and (d) show the radial intensity profiles of the corresponding color-separated images as black lines. For comparison to the first method, the dashed lines depict the respective contributions as obtained from fitting the total intensity profile with Mie solutions for a bichromatic field [compare Fig.~\ref{fig:full}]. The shaded areas are in the same position as in Fig.~\ref{fig:full}; they also mark where the pixel-based separation method fails, as evidenced by the experimental data deviating from the fits (dashed lines).
}
\label{fig:CDI2}
\end{figure}

Two-color diffraction patterns of pure HNDs were recorded for various experimental parameters, notably different intensity ratios of pump and probe pulse intensities and time delays between the two pulses, as well as varying HND sizes (Supplementary Fig. 2). A typical experimental diffraction pattern of a pure HND is shown in Fig.~\ref{fig:full}a. 
As inelastic contributions to the diffraction signal are negligible under the given experimental conditions, the image consists of a superposition of two patterns arising from the elastic scattering of the two X-ray pulses by the droplet. The nominal intensity ratio between the two colors is 1 to 4.5 ($4.2\times10^{15}$ to $1.9\times10^{16} \: $W/cm$^2$ based on beamline diagnostics~\cite{laksman2019, grunert2019}). The pump-probe delay in this example is set to 750~fs.
Since the European XFEL operates in the self-amplified spontaneous emission (SASE) mode~\cite{ayvazyan2002, altarelli2006}, the photon energies slightly vary from shot to shot (1\% of mean) around the mean value of $\SI{992}{\electronvolt}$ for the pump pulse and $\SI{1192}{\electronvolt}$ for the probe pulse. For simplicity, from now on we will call them ``red'' and ``blue'' pulses, respectively. To better visualize the image structure in the outer, sparsely illuminated regions of the detector, the radial intensity profile (black line) is added as an inset in the first quadrant of Fig.~\ref{fig:full}a. It is obtained by angular (azimuthal) averaging based on the count of non-masked pixels within each radial bin. In this way, masked pixels do not bias the radial intensity, though their exclusion slightly reduces the signal-to-noise ratio. Note that the image and the radial profile in the inset share the same radial scale.

Owing to the near-perfect spherical symmetry of HNDs and the absence of observable structural changes under the present experimental conditions, the radial intensity distributions of the measured diffraction patterns can be fitted by Mie's solution of Maxwell's equations for scattering by a hard sphere~\cite{Mie1908, bohren2004}. Molecular dynamics simulations, as further detailed in the supplementary materials, corroborate the latter assumption: due to the low X-ray absorption cross-section of helium at the employed photon energies~\cite{marr1976absolute}, the pump pulse initially photoionizes only 0.1\% of the atoms. Consequently, any structural dynamics occurring within the first picosecond are unobservable at the spatial resolution of our experiment.

Therefore, we can fit the radial profiles obtained from two-color scattering patterns by the sum of two simulated profiles, assuming a perfect sphere of unknown size and arbitrary contributions of both colors (see Methods).
Fig.~\ref{fig:full}b is a magnified view of the radial profile (black line) shown in the inset of Fig.~\ref{fig:full}a, including a fit of the radial intensity distribution of a bichromatic field scattered off a hard sphere~\cite{liu2007,colombo2024} (pink dashed line). The solid red and blue lines represent the individual color contributions that add up to the bichromatic fit. More examples of such radial profiles and fits are given in Supplementary Figs. 3 and 4.
The fit shows excellent agreement with the experimental data up to a scattering angle of $\theta \approx 1.5\,^\circ$, where the scattering intensity approaches the noise level. The decaying fringe pattern clearly shows a beat note arising from the superposition of fringes created by the blue and red pulses, where the fringe spacing differs by the ratio of photon energies, i.e., a factor of 1.2. The good agreement between the experimental data and the fit allows us to accurately quantify the HND's size and the absolute contributions of the two colors to the total diffraction pattern. We also tested the fitting of the time-evolved profiles with smoothed-surface model shapes. The results, however, showed no preference for significant smoothing at longer time delays, remaining consistent with hard edges within the resolution limit of our experiment.

While the comparison to Mie simulations allows us to separate the two color components of the radial intensity distributions in the specific case of spherical targets, it is desirable to develop a generally applicable approach for separating the two-color two-dimensional diffraction patterns. Therefore, we developed a pixel-based separation method based on the analysis of individual pixel counts and pattern identification. The amount of charge on the detector created by the photoelectric effect scales with the photon energy; the pnCCD records these photon-induced signals in analog-to-digital units (ADU). By decomposing each pixel ADU value into multiples of the corresponding charges, one can assign an active pixel to a unique combination of those photons. This method relies on the detector's high sensitivity to photon energy and low readout noise. An intrinsic challenge is the spillover of charges across pixel boundaries, creating clusters of hit pixels. Further details are described in the Methods.

\begin{figure*}
    \centering
     \includegraphics[width=\textwidth]{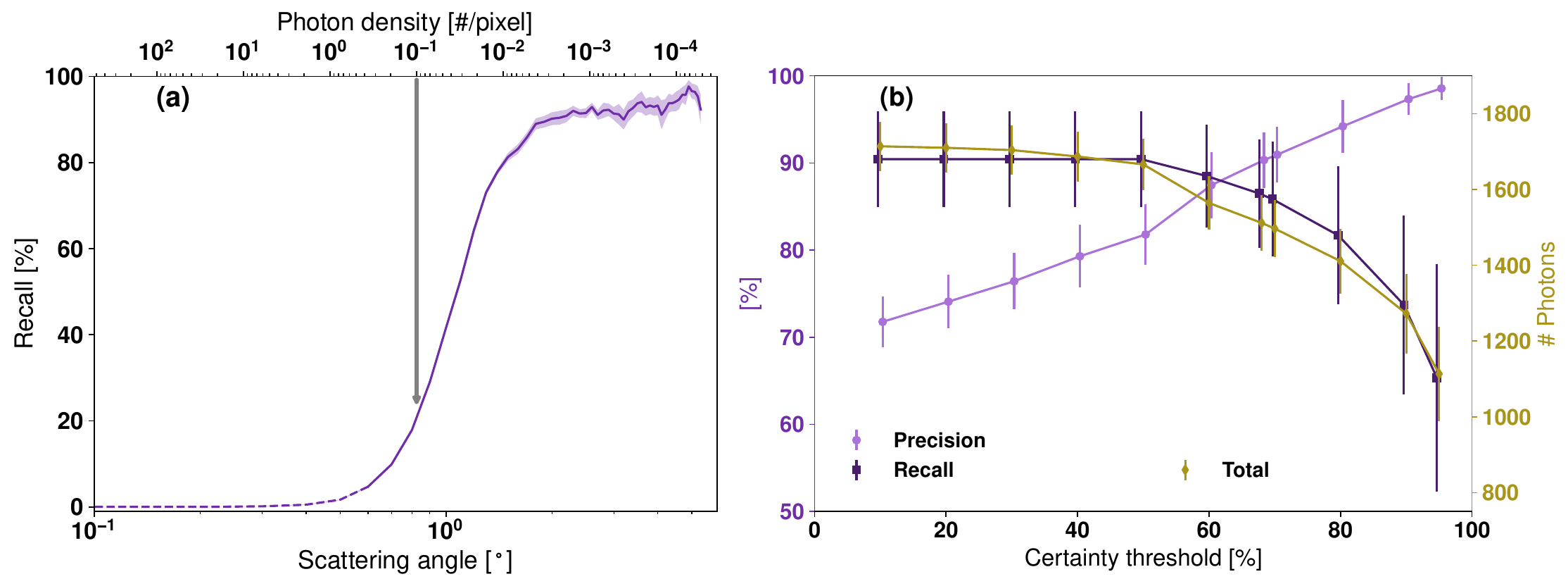}
    
    \caption{(a) Local recall of red photons over the scattering angle (lower axis) and the photon density (upper axis). The local recall is calculated from a set of synthetic images that have the same parameters as the experimental image in Fig.~\ref{fig:full}. The shaded band indicates the standard error of the mean. A steep rise around $0.8\,^\circ$ or $0.1$ photons per pixel marks the transition to a well-functioning pixel-based separation. The dashed part of the curve at angles $<0.55\,^\circ$ indicates the area omitted from the experimental data due to the central hole in the detector. The increase in uncertainty toward higher scattering angles $\gtrsim 2\,^\circ$, indicated by the shaded area around the curve, is attributed to the higher statistical fluctuations due to the lower photon count in that region. (b) Recall and precision of the pixel assignment for both colors, shown as a function of the certainty threshold. Error bars indicate the standard deviation. Recall is calculated cumulatively in the outer region of the detector ($>2\,^\circ$), where separation is limited only by ambiguities in color assignment, rather than by overlapping pixel clusters. Precision is evaluated over the entire detector area, as it considers only the pixels that were actually assigned. The right axis displays the total number of separated photons (in yellow). As this value is computed over the full detector, it shows slight deviations from the recall curve, which is presented in relative terms.}
    \label{fig:recall_cert}
\end{figure*}

Figs.~\ref{fig:CDI2}a and b show the result for the separation of the original diffraction pattern, Fig.~\ref{fig:full}a, into its red and blue components, respectively. The separated components' radial intensity profiles are presented in Figs.~\ref{fig:CDI2}c and d, respectively.
Clearly, the number of assigned pixels is significantly reduced in the color-separated images. First, signal is discarded if classified as noise or if the assignment to integer multiples of the red or blue elementary ADU values is ambiguous. Second, clusters of hit pixels are collapsed into the pixel position closest to the cluster's center of mass (see Methods for details). To quantify the level of ambiguity, we introduce a threshold parameter for the confidence in assigning an ADU value to either noise or a particular combination of red and blue photons. In this work, we choose a threshold of $50\%$ for the most likely photon combination (further details in Methods). For this value, the majority of identified pixel groups correspond to single-photon events, and $\sim10$\% of the color-separated photon events are attributed to two- or three-photon hits.

This identification of color contributions by pixel count is most effective in the outer areas of the image, i.e., at scattering angles of $\theta\gtrsim 0.8\,^\circ$ where lit pixels are sparse. 
At smaller angles, the probability for larger lit pixel clusters is higher, and pixel assignment becomes ambiguous. Thus, close to the center, intensity profiles generated from the color-separated images drop off, showing a significant signal loss. However, at high scattering angles, the sum of intensities for both red and blue pixel-based separated patterns presented in Figs.~\ref{fig:CDI2}a and \ref{fig:CDI2}b amounts to 87\% of the intensity for the original image in Fig.~\ref{fig:full}a. Further examples of similarly successful single-color pattern retrievals are given in Supplementary Figs. 3 and 4.%\ref{fig:exempImg} and \ref{fig:exempImg2}%

To test the limitations of this method and to provide boundaries within which it is reliable, we systematically generate synthetic diffraction patterns (see Methods) and color-separate them using the same algorithm used for processing the experimental images. This procedure allows us to benchmark the pixel-based image separation quantitatively, since, unlike real experimental data, the true photon distribution is known. To determine a threshold value for reliable pixel-based color assignment, we compute \textit{recall} and \textit{precision}, which are standard performance metrics for classification tasks~\cite{rijsbergen1981}. In our case, recall quantifies the ratio between the number of correctly assigned photons and the total number of photons. Precision measures the ratio of correctly assigned photons to all assigned photons, i.e., after discarding the ambiguous pixels. Figure~\ref{fig:recall_cert}a displays the recall as a function of scattering angle, calculated from a set of eleven synthetically generated patterns with radial profiles identical to those of the measured diffraction pattern in Fig.~\ref{fig:full}. The steep rise at an average photon density of $0.1$ detected photons per pixel indicates that efficient separation is feasible and reliable up to this threshold.

\section*{Discussion}
Our results demonstrate the feasibility of separating the initial-state diffraction pattern and the time-evolved diffraction pattern of the same nanoscale target obtained from the superimposed pattern generated by two X-ray pulses of different photon energies via computational methods. Although two-color Mie fits are only applicable to spherical particles, the high-quality separation achieved with Mie fitting in this work is a notable result on its own. The fits unambiguously yield an accurate droplet size and the absolute intensities of both X-ray colors at the position of the HND. While we use the method primarily for benchmarking the pixel-based color assignment in this work, the valuable information gained from the Mie fits could be included in more sophisticated combined approaches or for studying, e.g., ultrafast changes of optical properties~\cite{zimmermann2025}, where the droplet's refractive index is introduced as a fit parameter.

In contrast, pixel-based color separation is independent of target shape, and can be applied directly either as a stand-alone method or as a key step in the analysis pipeline in future two-color X-ray pump-probe experiments. Under the experimental conditions of this study, the separation works best for scattering angles of $0.8\,^\circ$ and higher, or equivalently, for areas of the detector detecting on average 0.1 photons per pixel and less. 

A deeper discussion of our tests with simulated diffraction patterns helps deduce guidelines for generalizing to other cases. The pixel-based assignment quickly loses its applicability in regions with high photon counts, as detected at low scattering angles in our diffraction patterns; the recall, shown in Fig.~\ref{fig:recall_cert}a, exhibits a sharp increase at approximately $0.8\,^\circ$. The origin of this clear step is the exponentially dropping intensity of small-angle diffraction from a homogeneous particle. This cut-off angle for unambiguous pixel identification depends in practice on photon flux per X-ray pulse, the nanotarget's size and shape, and the detector's distance from the interaction region. 

The recall reaches $90\%$ in the outer regions, where it is no longer constrained by large contiguously illuminated areas or high photon counts per pixel. This trend is further supported by the local photon density indicated on the top axis in Fig.~\ref{fig:recall_cert}a. Specifically, when the photon density falls below one photon per nine pixels --- implying that, on average, no neighboring pixels are illuminated --- the recall begins to rise. As detailed further in the Supplementary Information, this local photon density serves as a universal parameter governing the effectiveness of the separation method.

The level of the recall plateau depends on the choice of the certainty threshold. Increasing the required certainty, i.e., enhancing the selectivity to ensure that assigned events are correct, leads to a decrease in recall, as shown in Fig.~\ref{fig:recall_cert}b. In this figure, recall is evaluated at scattering angles $>2\,^\circ$ to assess the performance of the separation method irrespective of the issue of overlapping pixel clusters. In this regime, the recall is influenced primarily by the spread of the charge over multiple pixels and the intrinsic width of the resulting electron signal, which in turn determines the spread in ADU values. The total number of photons in Fig.~\ref{fig:recall_cert}b (on the right axis in yellow) behaves similarly to the recall but deviates slightly, because, unlike the recall, it is calculated for the whole image. The precision, i.e., the fraction of correctly assigned photons among all assigned photons, is also calculated over the entire pattern and increases almost linearly from $70\%$ to $95\%$.

For the analysis in this work, we choose a certainty threshold of $50\%$. This value strikes a balance between recall and precision, and it marks the point at which recall starts to decrease. One might want to decide differently in other experiments, e.g., increase the probability for correct assignment of events by setting a higher threshold. For bright small-angle coherent diffraction patterns, as presented here, the accumulation of signal in contiguously illuminated central regions will always pose a challenge for pixel-based separation. In other types of diffraction experiments, e.g., containing Bragg peaks, it might be possible to also assign higher photon counts to pixels, as the areas with high signal are spatially confined and thus not overlapping on the detector for appropriate choices of photon energies.

The inability of the pixel-based separation method to separate the two patterns towards the center of the image results in the loss of information at low momentum transfer, which corresponds to large spatial scales in the real-space object. To disentangle overlapping contributions in this central region, additional prior knowledge about the sample is required, such as in our case, the assumption of spherical shape and negligible change of the HND by the first X-ray pulse, allowing us to fit with Mie scattering solutions. This will certainly pose a challenge when dynamic processes are expected to occur across extended regions of the sample, e.g., in studies of ultrafast changes in the scattering cross section of materials near resonant photon energies~\cite{ulmer2025, kuschel2025}. However, for smaller-scale changes in HNDs, such as surface softening or expansion~\cite{peltz2014, bacellarAnisotropicSurfaceBroadening2022}, the higher scattering angles provide the most relevant information. In this regime, the pixel-based separation method performs particularly well.

One pathway to improve experimental conditions for pixel-based color separation is to increase the spatial resolution of the detector; either the pixel density could be enhanced, or the distance between the interaction region and the detector could be increased (see Supplementary Fig.~5). Then, the same diffraction features will spread out over more pixels, enlarging the area covered by sparse events. Additionally, 
the method could be further improved by optimizing the photon energy selectivity for the two X-ray pulses, ensuring that one-, two-, and potentially three-photon events are well separated across all combinations of two-color photons (see Supplementary Fig.~6 for various combinations). In this context, improving the signal-to-noise ratio of the detector will be crucial, too. %ref{fig:difPhoEne} is 6, 

In future experiments, the presented method has the potential to become a key building block for the analysis of a broad range of sophisticated two-color CDI approaches. By deliberately initiating structural dynamics with a first X-ray pulse while simultaneously recording the initial state and probing the system at a controlled time delay with a second X-ray pulse, the reconstruction of a time-resolved sequence and, ultimately, the creation of a true structural-dynamics ``movie'' of non-repetitive nanoscale objects in the gas phase would become possible. This approach could enable mapping of X-ray-induced ultrafast dynamics across a broad class of nanoscale systems, including pristine and doped helium nanodroplets, core–shell rare-gas clusters, nanofiber objects of various configurations, and complex biomolecules. By tuning photon energies across absorption edges, one could locally induce high degrees of ionization within a composite nanotarget, such as a helium nanodroplet with an embedded dopant nanostructure, thereby driving structural dynamics in a controlled way. The excitation of (inner-shell) electronic resonances also enables the study of ultrafast electron dynamics. Recent works indicate significant changes in X-ray responses through the excitation of transient resonances when few- and sub-femtosecond pulses are used~\cite{kuschel2025, ulmer2025}. By combining two-color CDI with such short two-color X-ray pulses, these intricate electron dynamics can be resolved in time and space.

While inelastic scattering does not affect the measured diffraction patterns in the current work, under resonant excitation or in strongly driven systems, nonlinear population dynamics, fluorescence, stimulated emission, or other inelastic contributions may become significant~\cite{kuschel2025, radloff2025, kruse2020}. Investigating such phenomena in single-nanostructures together with CDI could be a direct application of our pixel-based approach, given that the two signals are sufficiently separated in energy. In case of inelastic contributions occurring in a two-color experiment, conceptually corresponding to the addition of a third color, the algorithmic implementation in the pixel-based method would be straightforward.

Another intriguing approach is to induce dynamics with an additional intense optical laser pulse, while still capturing the particle's initial state with a preceding X-ray pulse. As long as the perturbations introduced by the first X-ray pulse are small, as is the case in the present experiment, the optical laser-induced dynamics would dominate and imaging of laser-driven dynamics such as melting and nanoplasma expansion in individual nanotargets would become feasible~\cite{gorkhoverFemtosecondAN2016,fluckigerTimeresolvedXrayImaging2016,langbehnDiffractionImagingLight2022,bacellarAnisotropicSurfaceBroadening2022,doldMeltingBubblelikeExpansion2025}.

\begin{figure*}[!ht]
	\centering
    \includegraphics[width=1.0\textwidth]{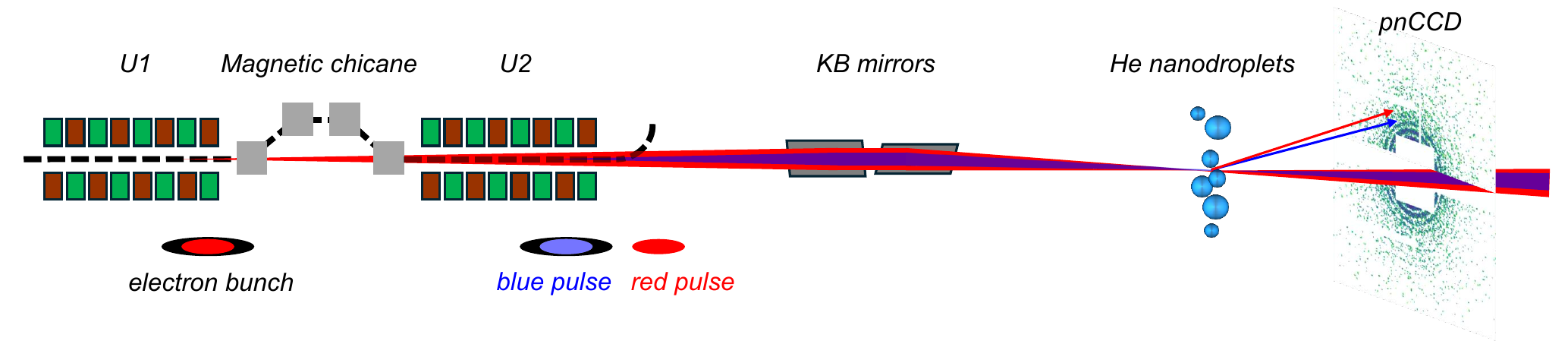}\caption{\label{fig:Setup} Schematic of the experimental setup used in this work. Two consecutive undulator sections (U1 and U2) are separated by a magnetic chicane. Because the chicane delays the electron bunch (black oval) relative to the first, red X-ray pulse generated in U1 (red oval), this pulse propagates ahead of the blue X-ray pulse subsequently generated in U2 (blue oval). The two X-ray beams are focused onto a jet of helium nanodroplets, and the diffraction patterns are recorded on an energy-resolving pnCCD pixel detector. See text and the Supplemental Material for a detailed description of the image-processing protocol.}
\end{figure*}

The clean wavelength separation at high-magnitude scattering vectors $ q $ will strongly benefit two-color imaging methods~\cite{hecht2026}, as it can provide reliable boundary conditions in pattern regions where phase-retrieval methods lack accuracy, helping determine accurate color ratios and, ultimately, enabling time-dependent structure determination within two-color phase-retrieval frameworks.

In summary, we presented a method for analyzing two-color coherent diffraction patterns. Having this versatile tool at hand to separate the diffraction signals from two-color X-ray pulses without requiring prior model assumptions about the initial state paves the way for recording genuine ultrafast X-ray movies with unprecedented temporal and spatial resolution.

\section*{Methods}

\subsection*{Experimental setup}
The experiment was carried out at the SASE3 soft X-ray branch of the EuXFEL~\cite{decking2020mhz} using the Nano-sized Quantum System (NQS) endstation at the Small Quantum Systems (SQS) instrument~\cite{SQS}. The EuXFEL was operated at an electron energy of $\SI{16.3}{\giga\electronvolt}$ with a pulse repetition rate of $\SI{10}{\hertz}$. Pairs of X-ray pulses with mean photon energies of $\SI{992}{\electronvolt}$ and $\SI{1192}{\electronvolt}$ were generated by simultaneously operating undulator sections U1 and U2~\cite{serkez2020opportunities}, see the schematic representation of the setup in Fig.~\ref{fig:Setup}. A magnetic chicane positioned between undulator sections U1 and U2 enabled control of the time delay between the two X-ray pulses (FWHM durations of 15–20~fs), allowing delays in the range of 50 to 750~fs. The two X-ray beams were focused by a Kirkpatrick–Baez mirror system to two spots with diameters of $\SI{5}{\um}$ (FWHM), separated by approximately $\SI{60}{\milli\meter}$ around the interaction point~\cite{serkez2020opportunities,mazza2023}. This displacement arises from the distinct source points of the two X-ray beams in the U1 and U2 undulator sections. To ensure approximately equal intensity ratios of the two X-ray pulses across the interaction region, the HND jet was aligned to intersect the X-ray beams at right angles midway between the two focal spots. Thus, the effective size of the X-ray spot is estimated to be about $\SI{14}{\um} $ FWHM. 

Due to slight fluctuations of the HND positions within the interaction volume, the X-ray intensity ratio of the two-color pulses in the diffraction pattern varied strongly from shot to shot. Many images even show only one color, as determined from the analysis of pixel values. In Fig.~\ref{fig:full}, for example, the nominal energies of the red and blue pulses were $0.16$ and $\SI{0.73}{\milli\joule}$, respectively,  as measured with a gas monitor in the tunnel, taking into account the beamline transmission of $0.6$. This corresponds to a nominal intensity ratio of $1:4.5$. In contrast, analysis of areas in the pixel-based separated diffraction patterns where recall is high revealed an intensity ratio of $2:3$, indicating that the HND was located closer to the red focus than to the blue focus in this specific event; see Supplementary Fig.~7.

We note that only elastically scattered photons were considered in our analysis. The inelastic scattering cross section of light elements around 1~keV, being far from resonance, is several orders of magnitude lower than the elastic contribution, especially in the forward direction~\cite{hau-riege2011}.

\begin{figure*}[!ht]
    \centering
    \includegraphics[height=0.27\textheight]{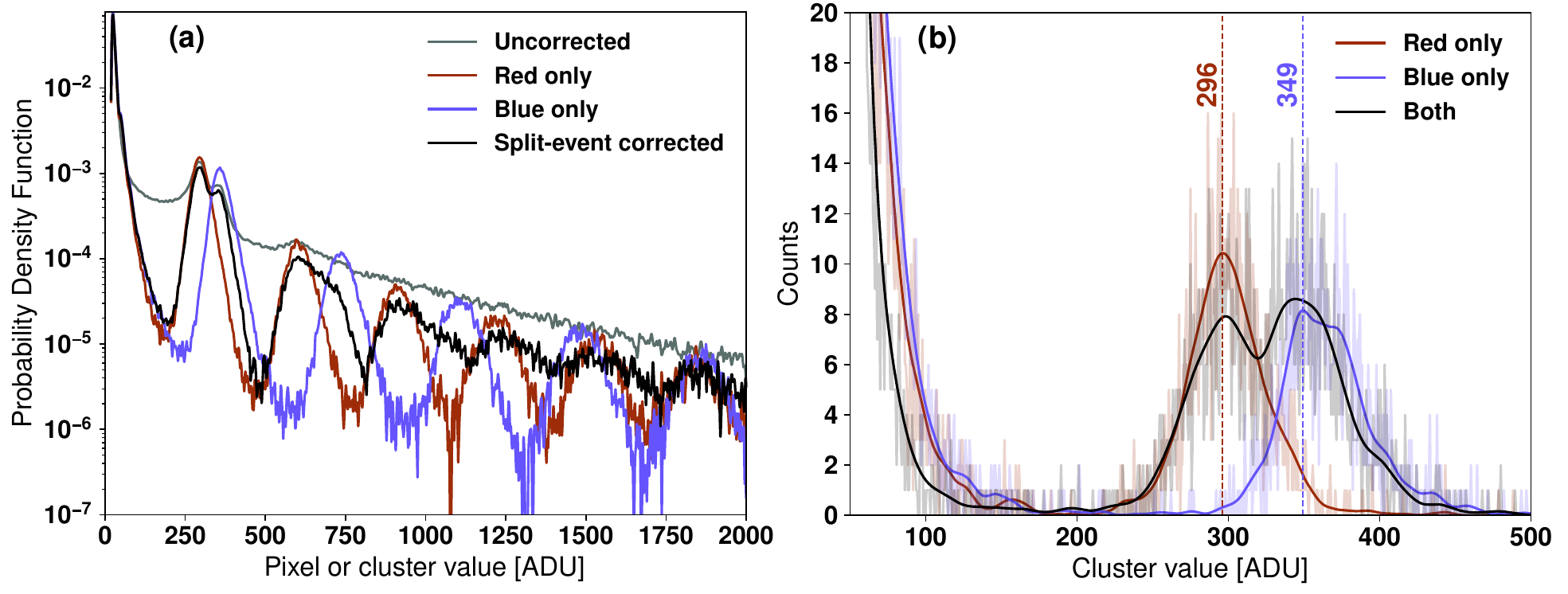}

    \caption{(a) ADU distributions for the 25 brightest two-color images, alongside single-color reference data. Distributions are shown as probability density functions (PDFs). The two-color data are shown before (gray) and after (black) split-event correction. Single-color data (red and blue) were recorded separately with undulators configured to a single discrete photon energy, each with comparable pulse energies; these are shown after split-event correction.
    (b) Close-up of single-photon events after split-event correction. Histograms (semi-transparent steps) and rescaled PDFs (solid lines) are shown for three individual shots: The two-color pattern from Fig.~\ref{fig:full}a (black), a single-shot red pattern (red), and a single-shot blue pattern (blue). Unlike the ensemble average in (a), the signal here is shown as individual counts. Single-color images are selected to have photon counts comparable to those of the two-color pattern. The vertical lines indicate the mean ADU values calculated from the full single-color datasets.}
\label{fig:histogram}
\end{figure*}

The HND jet was produced by an Even-Lavie-type cryogenic pulsed valve~\cite{pentlehner2009rapidly} operated using ultrapure helium gas (6.0) at a nozzle temperature of $ T_1=\SI{5}{K}$ or $T_2=\SI{10}{K}$ and at $p_1=\SI{50}{\bar}$ or $p_2=\SI{80}{\bar}$ stagnation pressure in two series of acquisitions. The resulting mean HND radii are estimated as $\overline{R_1}=\SI{370}{\nano\meter}$ and $\overline{R_2}=\SI{297}{\nano\meter}$, see Supplementary Fig.~2 for the whole size distributions. Under these conditions, the average rates for detecting a bright diffraction pattern were $0.3\%$ and $3.2\%$ per pulse, respectively.
The rate of detecting diffraction patterns of spherical HNDs illuminated by two X-ray pulses with comparable intensities (intensity ratio $\leq 3$) was 0.01\% and 0.26\%, corresponding to a total of $39$ and $93$ bright images for the two respective data sets. 

A large-area pnCCD detector (PNSensor GmbH)~\cite{kuster20211} was used for recording the two-color diffraction patterns. This detector is capable of counting single photons while offering a high dynamic range and low readout noise. This unique combination of detector properties is crucial for the effectiveness of the pixel-based separation method described in the following section. The pnCCD detector comprises two movable modules that can be separated to create a central gap, allowing the X-ray beam to pass through. The distance between the detector plane and the interaction region is adjustable and was set to 340~mm in the present experiment.

\begin{figure*}
    \centering
    \includegraphics[width=0.95\textwidth]{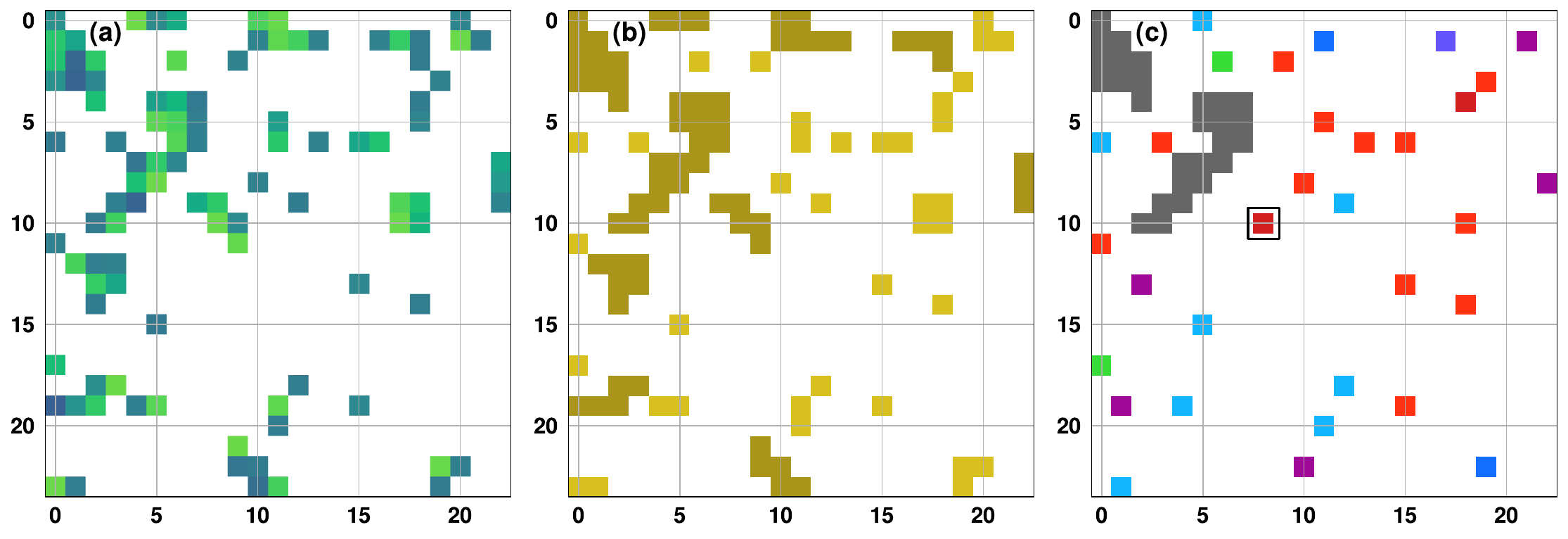}
        \caption{(a) Close-up view of the red boxed region in Fig.~1a. Pixel colors correspond to the measured ADU values according to the color scale in Fig.~1a. Relative pixel coordinates are indicated on the axes. (b) Result of the cluster-finding algorithm. Light yellow clusters correspond to standard patterns produced by single-photon events, while dark yellow clusters represent non-standard patterns likely resulting from multiple photons. (c) Final assignment of clusters to red and blue photon events based on total ADU values. Darker shades indicate two photons of the same color. Violet pixels denote mixed-photon events, green pixels identify detector noise, and unassigned clusters are shown in dark gray. The dark red pixel at (8,10), outlined in black, illustrates a two-photon event collapsed into a single pixel.}
    \label{fig:Closeup}
\end{figure*}
\subsection*{Pixel-based separation of images}
The pixel-based color-separation method exploits the pnCCD detector's ability to function as an X-ray imaging spectrometer, offering an energy resolution of $145$~eV (FWHM) at a photon energy of $1.4$~keV with a detector gain of $1/64$ ~\cite{kuster20211}.
Before applying the separation method to individual CDI frames, several correction procedures are performed, including background subtraction, masking of defective pixels, noise gating, common-mode correction, corrections for variable gain values, and charge transfer efficiency correction (see Supplementary Information for a detailed description). These corrections were applied to a set of the 25 brightest two-color diffraction patterns, and the resulting average distribution of analog-to-digital unit (ADU) values is shown as the dark green line in Fig.~\ref{fig:histogram}a. Here, the distribution is broadened due to the charge splitting of a single photon event across multiple detector pixels. By applying the split-event correction procedure (detailed below and in ~\cite{setoodehnia2020}), it becomes possible to distinguish single-, double-, and multiple-photon events, as illustrated by the black line in Fig.~\ref{fig:histogram}a. Clearly, the histogram of split-event corrected results is dominated by noise for ADU values up to approximately 200. The double-peak structure around 300 ADU corresponds to single-photon events from each color, while the broad maximum between 500 and 770 ADU arises from pixels where two-photon events generate overlapping charge clouds, resulting in ADU values approximately twice those of single-photon events. As the number of photons of each color increases within the same detector area, the ADU peaks in the histogram broaden roughly in proportion to the square root of the total ADU intensity (see Supplementary Fig.~8b), which in turn diminishes the ability to separate the two colors accurately. Simultaneously, averaged over the whole diffraction pattern, events involving two or more photons contributing to a single pixel count are approximately ten times less probable than single-photon events. To validate the color assignment indicated by the black line in Fig. 5 for the two-color mode, reference data were collected for each color individually under comparable experimental conditions, represented by the red and blue lines in the figure. 

A close-up view of the ADU distribution up to 500 for the single-shot two-color split-event corrected diffraction pattern, shown in Fig.~\ref{fig:full}, reveals a pronounced double-peak structure in the range of 200--400 ADU (black line in Fig.~\ref{fig:histogram}b). This structure is due to single-photon events of both the red and the blue colors, reflecting the photon energy distributions of the colors, which are distorted (or superimposed) by the pnCCD detector's 
instrument function. The assignment of the peaks in the ADU distributions to each color is confirmed by comparing to ADU distributions from single-color diffraction patterns.
This step is essential, as it determines the mean ADU values corresponding to single-photon detection for each individual color, which are subsequently used in the pixel-based separation method as reference values.

The pixel-based separation procedure of the measured total diffraction pattern into two individual components produced by the red and blue X-ray pulses is illustrated in Fig.~\ref{fig:Closeup}. Panel \ref{fig:Closeup}a provides a close-up view of the marked red area in Fig.~\ref{fig:full}a. 
Illuminated pixels are shown here in a gradient of green, with darker pixels corresponding to higher ADU values, indicating enhanced charge generation within the pixel. As the first step, our routine identifies all connected exposed pixels and assigns them to distinct clusters 
(see Fig.~\ref{fig:Closeup}b). 
Since a photon-generated charge cloud may be distributed across up to four neighboring pixels depending on its point of origin (see Supplementary Fig.~9,~\cite{setoodehnia2020}), in the second step, we identify these events and categorize them according to standard patterns (see light yellow pixels in Fig.~\ref{fig:Closeup}b). If more than one photon contributes to a pixel cluster, it is not assigned to contributing standard patterns and is marked as dark yellow pixels in Fig.~\ref{fig:Closeup}b. 
Nevertheless, in certain specific cases, such as the pixel at (8,10) in Fig.~\ref{fig:Closeup}c, which most likely originates from two photons of the same color, combinatorial reasoning can, in principle, uniquely trace a non-standard pattern back to the underlying single-photon patterns. We chose to omit this assignment step to maintain algorithmic simplicity. This results in a slightly reduced pixel-assignment precision, but it does not impact the overall spatial resolution of subsequent imaging algorithms. The resulting average shift of below one pixel remains well below the information limit defined by the diffraction fringe width and the intrinsic coordinate jitter from the FEL bandwidth. Furthermore, the sensitivity of real-space features to absolute pixel errors scales as $q^{-2}$, ensuring that the high-resolution information at high-q is inherently robust to minor positional uncertainties. We note that while this shift is negligible for oversampled, isolated-sample CDI, its impact would be more significant in experiments sensitive to absolute peak positions, such as Bragg diffraction or crystallography (see Supplementary Information for a detailed quantification).

Next, the entire pixel cluster is replaced by the single pixel closest to the cluster's center-of-mass, with its value set to the sum of all ADUs within that cluster. Finally, each pixel in the resulting map is categorized as detector noise (green), a single red or blue photon, or a mixed-color multi-photon event (see Fig.~\ref{fig:Closeup}). The accumulated charge, representing the photons per color, is indicated equivalent to Fig.~\ref{fig:full}a (compare colorbar).
Due to the overlap of the finite-width single-photon ADU distributions (see Fig.~\ref{fig:histogram}b), photon combinations can lead to ambiguous assignments. Although separating single or double photons remains highly reliable, resolving individual colors unambiguously becomes increasingly difficult at higher photon counts.

To reduce the risk of assigning incorrect combinations of red and blue photons, we apply a certainty threshold in the pixel-based separation method. This threshold defines a minimum confidence level required for photon assignment at each pixel. If a pixel's total ADU value falls within a region of ambiguity, where the likelihood of a correct match is below threshold, the cluster is excluded from further analysis by setting its value to zero, as illustrated by the gray areas in Fig.~\ref{fig:Closeup}c.

To compute the certainty threshold, we evaluate the probability that a given ADU value originates from a combination of photons, e.g., two red photons or one red and one blue photon. The combination with the highest probability is then assigned to the pixel. Each photon combination is modeled as a Gaussian intensity distribution, the width of which is determined by the single-photon peaks extracted from the averaged histogram for the brightest 25 shots (see Supplementary Fig.~8). The height and exact position of each Gaussian are obtained from fitting the histogram for each individual image. This procedure accounts for shot-to-shot intensity fluctuations caused by fluctuations of the position of the X-ray beams and the sample jet, as well as jitter in the central wavelength inherent to the SASE mode. The final certainty is then calculated as the relative contribution of the Gaussian fit with the highest amplitude (i.e., the assigned photon combination) to the sum of all contributing photon combinations at that ADU value. 

To determine the most probable photon combination for a multi-photon event, we model the expected occurrence of mixed-color clusters based on the global intensity ratio of the two colors. Rather than assuming a homogeneous flat-field distribution, where mixed events follow a simple binomial distribution, we account for the structured nature of the diffraction signal: the color-intensity ratio and thus the probability for each color vary as a function of scattering angle. This includes areas where predominantly red photons are expected and vice versa. The exact distribution depends on the specific scattering conditions. Here, the total intensity, the intensity ratio, and the nanoparticle's radius are defining parameters. Since the pixel-based method remains agnostic to the exact size and local scattering probabilities, we linearly scale the expected probability of each color combination based on the average intensity ratio across the entire detector as a reasonable default assumption for the average probabilities across the whole detector. For example, in the case of similar overall intensity of the two colors, the probability of a two-photon event of mixed colors is equal to that of two photons of the same color. This relationship is supported by both mathematical considerations and numerical tests on simulated data, as further discussed in the Supplementary Information.

Fig.~\ref{fig:Confidence} shows the individual contributions of different combinations of photons, along with the resulting certainty level indicated by a black line. For illustrative purposes, the same intensities as in the experimental image are chosen. As an example, we focus on the left wing of the ``one red photon peak'', around 250 ADU. As seen, the ``one red photon event'' dominates the likelihood distribution, contributing nearly unity. Therefore, assigning a cluster with an ADU value of 250 to a single red photon is unambiguous. In contrast, around 320 ADU, where the probabilities for red and blue photons are approximately equal, and each contributes less than one, the assignment becomes less certain. Clusters with ADU values in this ambiguous range cannot be reliably attributed to any specific photon color and should be excluded from further analysis when a high certainty threshold is required. At higher ADU values, near the ``one blue photon peak'' at 349 ADU, the likelihood is again dominated by a single contribution, making the assignment to a single blue photon event reliable.

\begin{figure}[tb]
   \centering
    \null\hfill
        \includegraphics[height=0.25\textheight]{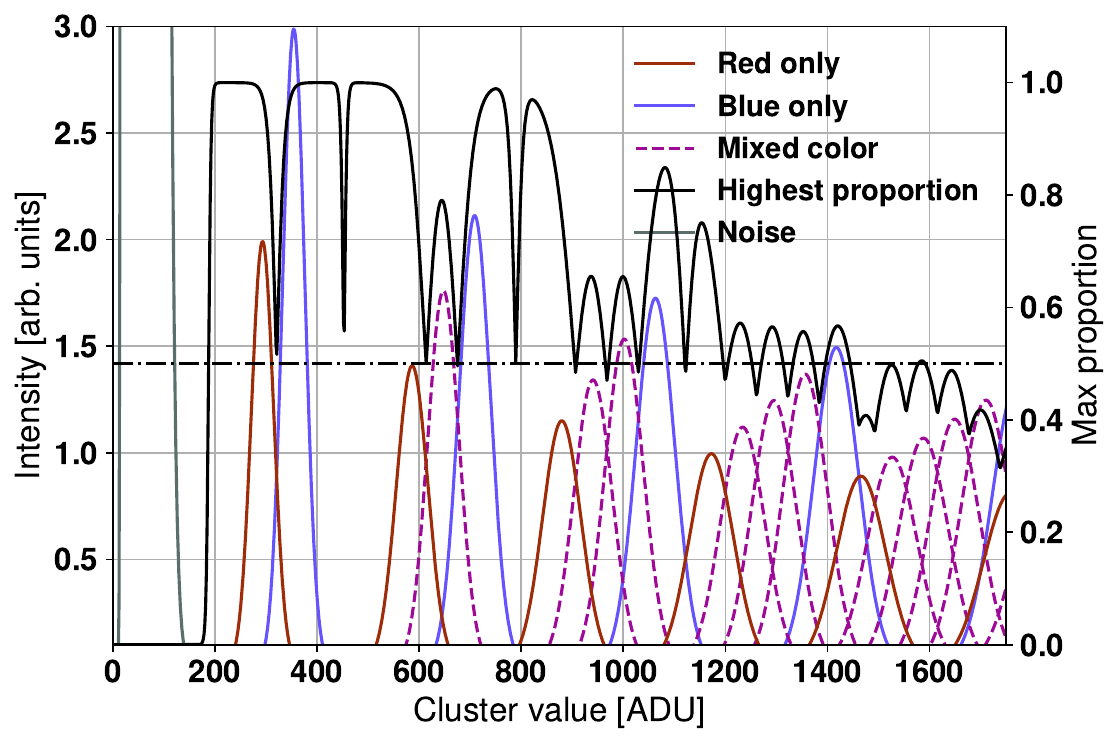}
    \hfill\null
    \caption{Calculated probability distributions for red, blue, and mixed photon events, shown as red, blue, and purple lines, respectively (referenced to the left ordinate). The gray line represents the contribution from pixel noise. The black line (referenced to the right ordinate) indicates the maximum proportion of each event type at a given ADU value. The horizontal line marks the certainty threshold set to 50\%, as discussed in the text.}
    \label{fig:Confidence}
\end{figure}

\subsection*{Synthetic image generation}

\begin{figure*}
    \centering
    \includegraphics[width=\linewidth]{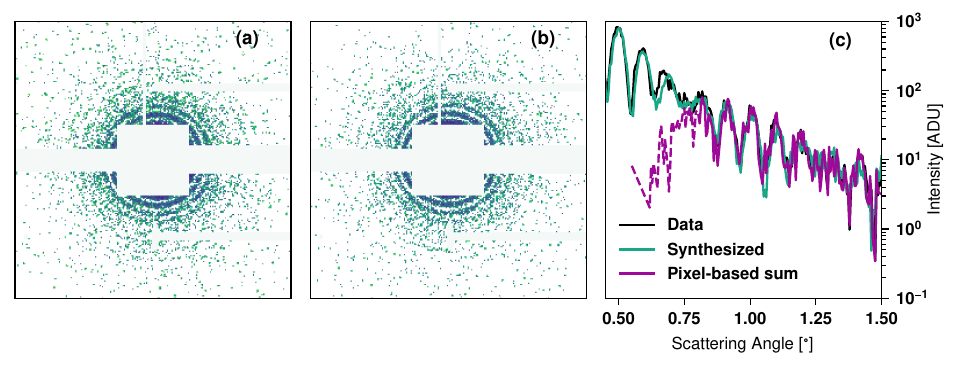}
    \caption{(a) Comparison between the experimentally measured two-color diffraction pattern from Fig.~\ref{fig:full}a and (b) a synthetic image generated for the same HND size and pulse intensity as in the experiment. Both share the color scale of Fig.~\ref{fig:full}. (c) Radial profiles of the measured image, the synthetic image, and the sum of the two images retrieved by pixel-based separation of the measured image.}
    \label{fig:comp_synth}
\end{figure*}

To evaluate the pixel-based separation method and to guide the selection of an appropriate certainty threshold, we generate synthetic image data that closely replicates the experimental results. 
We assume a static and spherical target and the experimental conditions described above.

The intensities of the red and blue images are determined from the components of the Mie fits to the radial intensity profiles of the experimental images (see next section). For both colors, the corresponding ADU values are fitted by Gaussian distributions using data from the 25 brightest images to infer the mean value and the spread. To incorporate photon statistics, the mean ADU values are used to assign the number of photons per pixel randomly according to a Poisson distribution. To account for detector noise, the ADU values for each photon are then drawn from Gaussian distributions based on their respective spreads. As explained earlier, each photon generates an electron cloud that can spread across up to four neighboring pixels, creating characteristic patterns (see Supplementary Fig.~9). We use the experimentally determined distribution of these pixel patterns to randomly assign a pattern to every photon event (Supplementary Table~1). For each pattern, the total ADU value is distributed evenly across the pattern's pixels, with the original pixel where the photon event was placed according to Poisson statistics receiving the highest value. The total diffraction pattern is obtained by summing all single-photon events of both colors and adding dark noise, as observed in the experimental images. In this process, stray light signal seen by the detector, as well as detector artifacts such as hot pixels and broken rows, are neglected because these are typically masked and excluded from analysis of the experimental data. 

Figure~\ref{fig:comp_synth} presents a direct comparison between an experimental pattern and a corresponding synthetic image. The synthetic image (b) closely reproduces the features of the original experimental image (a). The radial intensity profiles of both images, shown in panel (c), also compare favorably, demonstrating that the synthetic approach provides a reliable approximation of the experimental data.

\subsection*{Separation of images by Mie solutions}
Owing to the nearly ideal spherical shape of HNDs and the resulting radial symmetry in their diffraction patterns, an alternative method for separating the two color components of the diffraction patterns can be employed. The scattering of light from a spherical HND can be accurately described by Mie theory~\cite{Mie1908, bohren2004}.  
Here, we use Mie solutions to fit the radial intensity distributions of the measured diffraction patterns. 
Specifically, the optical properties of liquid helium are used~\cite{henkeXrayScatteringFactor1993}. We employ the Python library \textit{nux-utils}, which is based on C++ code~\cite{colombo2024, liu2007}. \textit{nux-utils} is used to calculate all radial intensity profiles by averaging over the azimuthal angle; the signal is normalized to the number of valid pixels per bin, while masked pixels are excluded.

Since the measured images result from the superpositions of two independent scattering events caused by red and blue X-ray pulses interacting with the same particle, we use three fit parameters: two for the intensities of each pulse and one for the size of the HND. The center of each individual diffraction pattern depends on the position of the HND relative to the wavefront of the incident X-ray pulse. To account for shot-to-shot jitter, the center of each image is determined by a fitting procedure before further analysis. Since Mie solutions assume perfectly spherical particles, a preselection of radially symmetric diffraction patterns is required. An example of the fit result is shown in Fig.~\ref{fig:full}b, where the black line represents the radial profile of the measured image. The red and blue lines correspond to the two components of the fit model, with their sum shown as a purple line. 
In the case of scattering from a spherical object with radius $R$ under the condition $qR\gg1$, which holds here with $qR=16$ for $R=300$~nm and $\theta = 0.5\,^\circ$, the scattered intensity falls off steeply with the magnitude of the scattering vector, following the relation $I(q)\propto q^{-4}$, where $q = \frac{2 \pi}{\lambda}\sin{\theta}$ denotes the magnitude of the scattering vector, $\lambda$ is the photon wavelength, and $\theta$ is the scattering angle between the incoming beam and the detected light on the detector. Therefore, the residual is scaled by $q^4$ to equally weight data points across the entire $q$-range considered in the fit. Additionally, the Fourier spectra of both the fit and the data are incorporated into the error function. This enhances the robustness of the fit procedure and improves the accuracy of HND size determination, particularly in the presence of noise.

\begin{figure*}
    \centering
    \includegraphics[width = \textwidth]{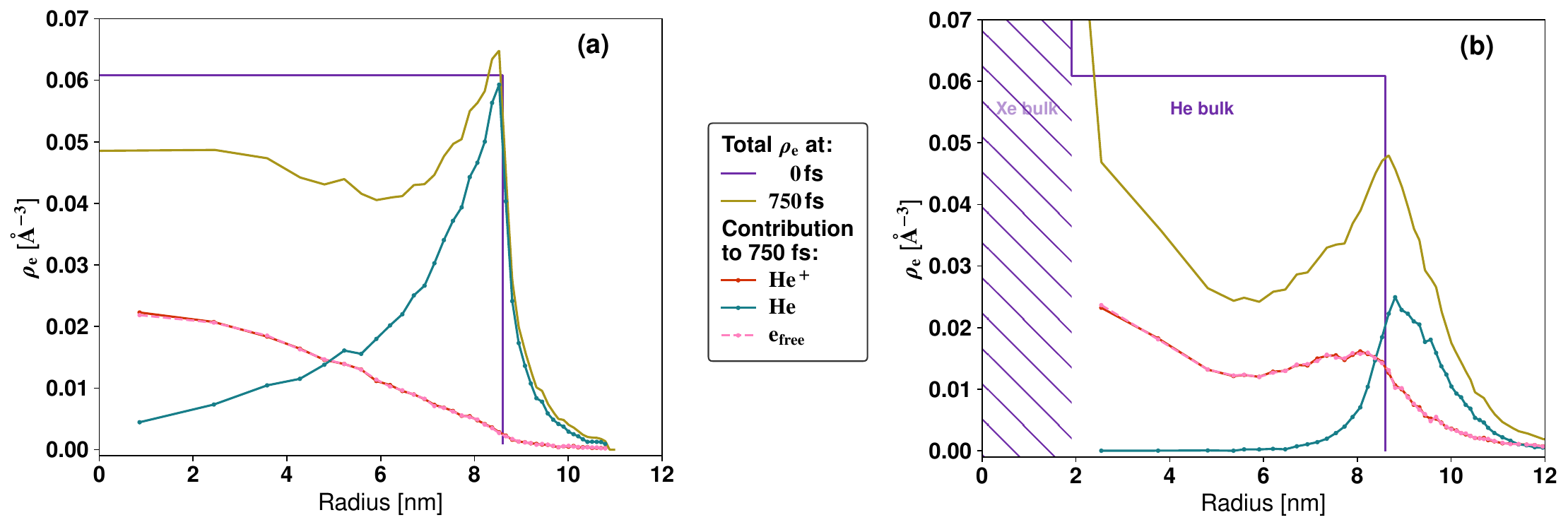}
    \caption{(a) Radial density distributions of neutral $\mathrm{He}$ atoms, $\mathrm{He}^+$ ions, quasi-free electrons $e$ and the total electron number density $\rho_{e,tot}$ at $t=750$~fs obtained from a classical MD simulation of the time evolution of a partially ionized HND containing $8\times 10^4$ atoms. At $t=0$, a fraction of 2\% of He atoms is photoionized by 1~keV photons. (b) Radial density distribution of the same-sized helium droplet doped with 500 xenon atoms in the center instead. Here, the irradiation at $t=0$ is set to be at the nominal experimental pump intensity of $4.2\times10^{15}\: $W/cm$^2$, a factor 100 lower than in (a). Still, due to the high energy absorption of the $\mathrm{Xe}$ atoms, the magnitude of the expansion, i.e., the smoothening of the surface, is roughly four times greater.}
    \label{fig:MD}
\end{figure*}

\subsection*{Test for dynamics by Mie fits}
To cross-check our assumption of static HNDs, we perform additional Mie fits. In these cases, we add an additional degree of freedom that mimics typical dynamics: The first possibility would be a softening of the droplet surface that reflects as a steeper drop of the radial profile toward larger scattering angles in the diffraction pattern~\cite{peltz2014}. This effect is modeled by weighting the simulated probe Mie profile with a Gaussian function peaked at zero. The second possibility would be an increase in droplet size. This can be easily accounted for by lifting the restriction that both Mie profiles must be for the same size of the particle. Both cases were tested, but no statistically relevant differences were found for any delay ($50$ and $750$~fs). We take this result as a confirmation of the lack of structural dynamics of the HNDs under the present experimental conditions, in agreement with the presented molecular dynamics simulations, see below. An upper limit for changes of HND size is given by the spatial resolution of the detector, $\pi /q = \SI{5}{\nano\meter}$, determined by the detector geometry and photon energy~\cite{bostedt2016}. However, due to the relatively low pulse energies in the two-color mode, the largest scattering angle up to which the signal-to-noise ratio is sufficient is significantly reduced to $\theta = 1.5\,^{\circ}$, resulting in an effective resolution of $\SI{20}{\nano\meter}$.

\subsection*{Molecular dynamics simulations}
To gain insight into the ionization dynamics of HNDs irradiated by X-ray pulses, we performed classical molecular dynamics (MD) simulations. A detailed description of the method can be found in Refs.~\cite{heidenreich2007control,heidenreich2009extreme,heidenreichEfficiencyDopantinducedIgnition2016,medinaSingleshotElectronImaging2021}. Due to computational constraints, the simulated HNDs were significantly smaller than those used in the experiment, reaching up to $8\times 10^4$ He atoms ($R=8.6$~nm).

Given the experimental photon fluence of approximately $ 4\times 10^{17}$~photons/cm$^2$ per pulse and the photoionization cross section of helium at 1~keV of $5.6\times 10^{-22}$~cm$^2$ ~\cite{marr1976absolute}, we estimate that only about 0.02\% of the He atoms in the HNDs are photoionized. At such a low degree of photoionization, no measurable structural change of the HNDs is expected within the timescale of the experiment (750~fs). To explore the intensity regime required for inducing detectable structural dynamics, the MD simulations for pristine He were performed at significantly higher fluences.

As an example, 2\% of the He atoms were photoionized by 1~keV photons, assuming a uniform probability distribution across the HND. Photoionization was modeled as an instantaneous, \textit{ad hoc} process at $t = 0$, and the MD trajectories were propagated up to $t = 750$~fs. Figure~\ref{fig:MD}a shows the radial distributions of the number density of electrons bound to He atoms, $\mathrm{He}^+$ ions, and quasi-free electrons, plotted up to a distance of 1.25 times the HND radius ($R=8.6$~nm). The radial density function of the HND at $t = 0$ is indicated by the purple curve. The highest $\mathrm{He}^+$ ion density (red curve) is observed near the droplet center. It is consistent with previous simulations for Xe clusters~\cite{heidenreich2007control,heidenreich2009extreme} which suggest that secondary electron impact ionization, triggered by initial photoionizations, occurs preferentially in the cluster interior. The curves for $\mathrm{e}_\mathrm{free}$ (pink) and $\mathrm{He}^+$ (red) almost perfectly overlap, indicating net charge neutrality across the interior and in the vicinity of the HND. The droplet periphery consists mainly of neutral He atoms (teal curve). The resulting $\rho_{e}(r)$ curve (orange) features a dent inside the HND, which spreads into the surrounding outer region of the HND with time as the HND ejects He atoms, ions, and electrons. 

Similar but more pronounced dynamical effects can be observed when placing a highly absorbing species into the droplet core. In Fig.~\ref{fig:MD}b, we studied 500 xenon atoms in a helium droplet of the same size. Since energy deposition in the xenon core drives more pronounced structural changes, we were able to use the actual experimental pump-pulse intensity of $4.2\times10^{15}\: $W/cm$^2$ in our simulations, accounting explicitly for the
Gaussian temporal pulse profile (FWHM $\SI{20}{\femto\second}$), the photoionization
cross sections of Xe ($\SI{2.1}{\mega\barn}$) and of He ($\SI{4e-4}{\mega\barn}$)~\cite{yeh1985}, and
Auger emissions of Xe~\cite{eland2015}. The overall smoothing of the surface follows a behavior similar to that in the pristine case. The expansion effects for such a small particle are roughly four times more pronounced, judging from the expansion at the half maximum point of the electron density.

Direct scaling of these observations to the 50 times larger droplets in the experiment is not straightforward. In both cases, the simulated response of the smaller particles provides only an upper bound for the structural expansion expected at the experimental particle size. Nevertheless, this estimate provides a clear indication that even in the 100-fold exaggerated case of 2\% ionization, the resulting size increase of less than 5\% for a $\SI{400}{\nano\meter}$ droplet would not be resolvable given our $\SI{20}{\nano\meter}$ resolution limit. For the purposes of benchmarking and cross-checking our methods, the lack of structural dynamics turned out to be beneficial, as it justifies the use of the two-color Mie fitting routine. The results from the more scientifically interesting scenario of helium droplets doped with a heavy-species structure suggest that the experimental observation of X-ray-induced dynamics using time-resolved single-particle CDI is within reach.

\section*{Data availability}

All raw data generated in this study have been deposited in the EuXFEL database under accession code doi:\nolinkurl{10.22003/XFEL.EU-DATA-002857-00}.

\section*{Code availability}
The Mie fitting routine is based on the package nux-utils: \url{https://gitlab.ethz.ch/nux/essential-tools/nux-utils}.
The pixel-based separation method is exemplified here: \url{https://gitlab.ethz.ch/nux/publications/pixel-based-separation}.

\section*{Acknowledgements}
A.H. gratefully acknowledges computational resources and manpower support provided by the DIPC computation center. The authors acknowledge EuXFEL in Schenefeld, Germany, for provision of X-ray free-electron laser beamtime at SQS SASE3 (proposal 2857) and would like to thank the staff for their assistance.

\section*{Funding Statement}
L.H., B.B., A.L., D.R., and M.Mu. thank the Danish Agency for Science and Higher Education for support through the instrument center DanScatt. A.H. gratefully acknowledges financial support from the Basque Government (project IT1584-22). S. K. is grateful for the support from the Department of Science and Technology (India-DESY and Core Research Grant) as well as Ministry of Education (Institute of Eminence Scheme) and high-risk-high-reward programme of IIT Madras. F.S. acknowledges funding by the DFG of the RTG DynCAM (RTG2717). T.F. and C.P. acknowledge funding by the DFG via CRC 1477 ``Light-Matter Interactions at Interfaces'' (project ID: 441234705). The research leading to these results was supported by SNSF project grant 10003697 and by the COST Action CA21101 ``Confined Molecular Systems: From a New Generation of Materials to the Stars (COSY)''. 

\section*{Author contributions statement}
M.Me., Y.O., M.Mu. conceived the experiment. L.H., Y.O., B.B., T.M.B., A.C., S.De., A.D.F., S.D., S.K., B.K., B.L., T.M., S.M., C.M., T.P., B.S., K.S., F.S., R.M.P.T., P.T., S.U., D.R., M.Mu. conducted the experiment. L.H., Y.O., A.L., A.C., R.H., T.F., C.P., A.H., M.Mu. analyzed the results. L.H., Y.O., A.L., T.F., C.P., K.K., M.Me., D.R., M.Mu. prepared the manuscript, with input from all authors. 

\section*{Competing Interests Statement}
The authors declare no competing interests. 
\end{document}

\title{Supplementary Information for: \linebreak Model-free pattern separation of two-color ultrafast X-ray diffraction}

\author{Linos \surname{Hecht$^{*}$}}\affiliation{\affila}\email{lhecht@phys.ethz.ch, yevheniy.ovcharenko@xfel.eu,\\ruppda@ethz.ch, mudrich@uni-kassel.de}
\author{Yevheniy \surname{Ovcharenko$^{*}$}}\affiliation{\affilb}
\author{Asbjørn Ø. \surname{Lægdsmand}}\affiliation{\affilc}
\author{Björn \surname{Bastian}}\affiliation{\affild}
\author{Thomas M. \surname{Baumann}}\affiliation{\affilb}
\author{Alessandro \surname{Colombo}}\affiliation{\affila}
\author{Subhendu \surname{De}}\affiliation{\affile}
\author{Alberto \surname{De Fanis}}\affiliation{\affilb}
\author{Simon \surname{Dold}}\affiliation{\affilb}
\author{Thomas \surname{Fennel}}\affiliation{\affilf}
\author{Robert \surname{Hartmann}}\affiliation{\affilg}
\author{Katharina \surname{Kolatzki}}\affiliation{\affila}
\author{Sivarama \surname{Krishnan}}\affiliation{\affile}
\author{Björn \surname{Kruse}}\affiliation{\affilf}
\author{Aaron C. \surname{Laforge}}\affiliation{\affilh}
\author{Bruno \surname{Langbehn}}\affiliation{\affili}
\author{Suddhasattwa \surname{Mandal}}\affiliation{\affilj}
\author{Tommaso \surname{Mazza}}\affiliation{\affilb}
\author{Cristian \surname{Medina}}\affiliation{\affilk}
\author{Christian \surname{Peltz}}\affiliation{\affilf}
\author{Thomas \surname{Pfeifer}}\affiliation{\affilk}
\author{Björn \surname{Senfftleben}}\affiliation{\affila}\affiliation{\affilb}
\author{Keshav \surname{Sishodia}}\affiliation{\affile}
\author{Frank \surname{Stienkemeier}}\affiliation{\affill}
\author{Rico Mayro P. \surname{Tanyag}}\affiliation{\affilo}
\author{Paul \surname{T\"ummler}}\affiliation{\affilf}
\author{Sergey \surname{Usenko}}\affiliation{\affilb}
\author{Andreas \surname{Heidenreich}}\affiliation{\affilm}\affiliation{\affiln}
\author{Michael \surname{Meyer}}\affiliation{\affilb}
\author{Daniela \surname{Rupp$^{*}$}}\affiliation{\affila}
\author{Marcel \surname{Mudrich$^{*}$}}\affiliation{\affilc}
\affiliation{\affilp}

\maketitle
\renewcommand{\thefigure}{\arabic{figure}}
\setcounter{figure}{0} 
\renewcommand{\figurename}{Supplementary Fig. }
\makeatletter
\renewcommand{\fnum@figure}[1]{Supplementary Fig.~\thefigure.~}
\makeatother

\renewcommand{\thetable}{\arabic{table}}
\setcounter{table}{0}
\renewcommand{\tablename}{TAB.}
\makeatletter
\renewcommand{\fnum@table}[1]{Supplementary Table~\thetable.~}
\makeatother

\phantomsection
\section*{}
\subsection*{Detector corrections}\label{detCor}
We thoroughly processed the diffraction patterns before applying the pixel-based separation method. How the detector works is described in detail in [38]. Below, the correction steps are listed and briefly described. The first two steps are applied to all data; all further steps are updated during the experiment:
\begin{enumerate}[itemsep=0.0pt, topsep=0.5pt,leftmargin=*]
    \item \textbf{Charge transfer efficiency} map: As the charge is read out at one of the two lateral edges, every pixel's charge needs to be transported to the outside. In doing this, some charge is lost during the transport process. Using the known charge transfer efficiency, the signal of the more central columns is increased accordingly.
    \item \textbf{Gain} map: Every detector row has an individual gain. Before the experiment, we created a homogeneous signal through fluorescence at a signal level where, on average, only single photons are sparsely distributed on the detector. Collecting row-wise all events with only a single pixel activated creates an ADU distribution for every row. The gain map multiplies every row so that the average ADU value of a single photon is the same for all rows.
    \item \textbf{Background} pattern: Regularly, we took background runs, meaning no scattering target was introduced into the interaction region. We take 20000 patterns and delete every pixel's brightest and darkest ten values. Afterward, the remaining values are averaged. This background pattern is then subtracted from the data.
    \item \textbf{Mask}: As the detector already has some damaged areas, we mask some parts. We use the mean and standard deviation of every pixel to determine which pixels to mask. Generally, we mask generously by choosing strict values and excluding surrounding pixels, as we want to support pixel-based separation as much as possible. We exclude a pixel if
    \begin{enumerate}[itemsep=0pt, topsep=0pt, ]
        \item the mean value is above a threshold (hot pixel).
        \item the standard deviation is too big (flickering pixel) or too small (stuck pixel).
        \item the ratio between mean and standard deviation is too big or too small.
    \end{enumerate}
    \item \textbf{Common-mode correction}. During the readout process of a single pattern, the electronics are subjected to noise. This results in a different baseline for every column of pixels. As this is constant for the entire column, it can be removed afterward by subtracting the median of each column, provided the signal does not cover most of the column.
    \item \textbf{Noise rejection}: We create a noise map from the background run. After background subtraction and common-mode correction of each pattern, we calculate the standard deviation of every pixel. If a pixel's value in the data is below twice the standard deviation, we mark it as detector noise and remove it.
    \item \textbf{Detector plate positioning}. The detector halves are movable in the detector plane and were moved to reduce stray light on the detector. Using spherical hits as a reference, we add pixels in the center of the pattern matrix and align the plates laterally by adding blank pixels at the edges.
    \item \textbf{Center finding}. The exact center of the scattering pattern fluctuates from shot to shot. It depends on the position of the HND in the pulse's bent wavefront. We fit a center position to every pattern using the point symmetry of the scattering pattern.
\end{enumerate}

\subsection*{Pixel-based cluster assignment}
The processed raw patterns are passed through a series of steps to classify the pulse components of different parts of the pattern. First, the clusters of bright pixels in each pattern are labeled, noting the total brightness and their position. The relative intensity of the red and blue pulses is estimated. Finally, each cluster is assigned photon components based on its ADU value.

Split events, where photons partly activate more than one neighboring pixel, comprise about a third of all single photon hits in flat illumination. The image detection of the clusters was done in two steps. The one-photon signal can only create specific 1-4 pixel clusters. These were first labeled by convolving the pattern with matrices chosen to capture each shape. Afterwards, the remaining clusters were identified using a standard connected component analysis algorithm in SciPy. 

Because each diffraction pattern can result from scattering off droplets anywhere between the two foci of the pulses, the resulting patterns contain different ratios of red and blue photons. Using the histogram of the cluster ADU values, similar to Fig.~5, the relative sizes of the one-photon and noise peaks are found. For these fits on individual patterns, the heights and the positions of the peaks are varied. The other parameters determining their shape were assumed to be the same for all patterns. %\ref{fig:histogram}is figure 5

We now consider the signal in each cluster as resulting from a single-photon or multi-photon event, with a mean value of a photon $\mu$ and a standard deviation $\sigma$. Under the assumption that the signal from each pump or probe photon is normally distributed ($N$), the resulting distribution $S_\text{n, m}$ of a given multiphoton event can be estimated.
\begin{align*}
	S_\text{pump} &\sim N(\mu_\text{pump}, \sigma_\text{pump}^2)\\
 	S_\text{probe} &\sim N(\mu_\text{probe}, \sigma_\text{probe}^2)\\
	S_{n,m} &\sim N(n\mu_\text{pump} + m\mu_\text{probe}, n\sigma_\text{pump}^2+m\sigma_\text{probe}^2) \\
    &\hspace{1cm}\text{for }n,m\in \mathbb{N}.
\end{align*}

\begin{figure*}[]
    \centering
    \null\hfill
    \includegraphics[width=.49\linewidth]{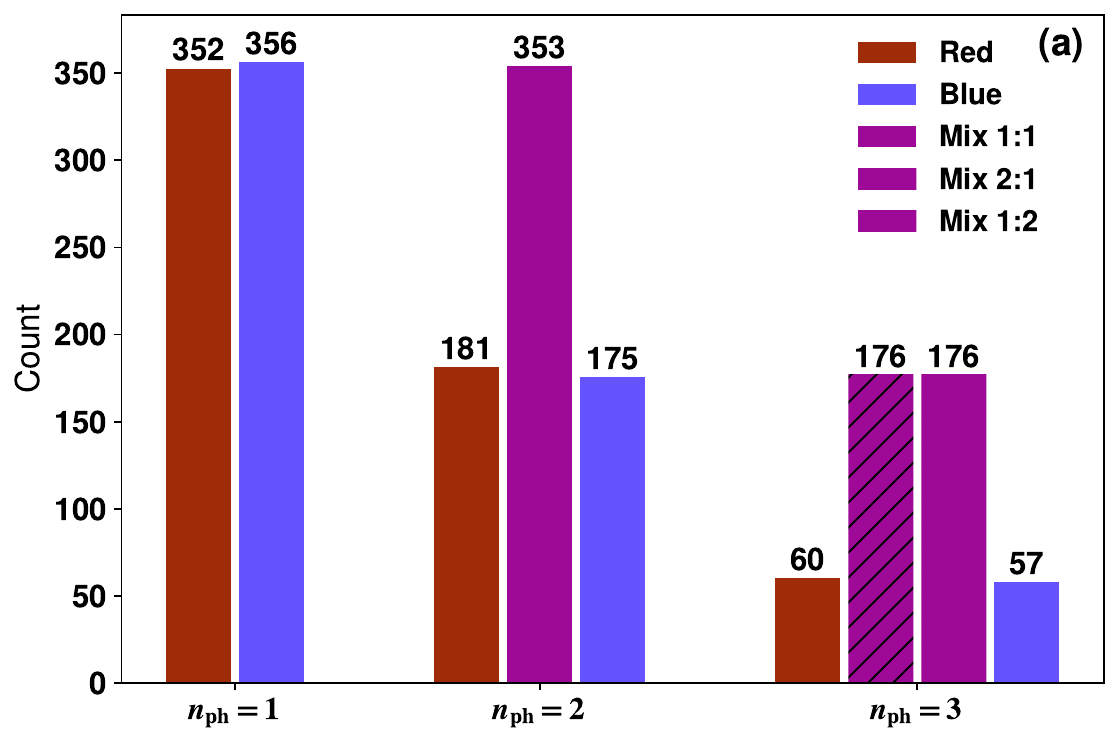}
    \hfill
    \includegraphics[width=.49\linewidth]{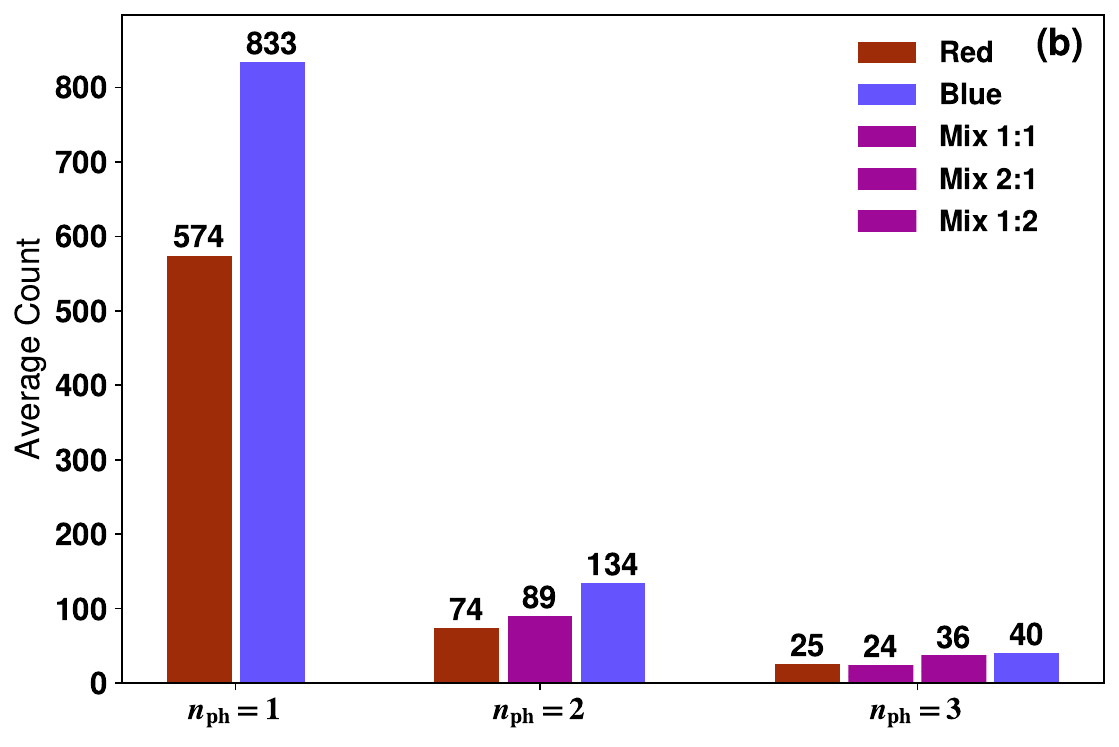}
    \hfill\null
    \caption{Histogram of photon counts per pixel on simulated data using the experimental photon energies. The number of photons in a pixel with all possible color combinations is shown as bars (see legend). (a) a flat field with an average photon density of 1. It reproduces the binomial distribution. (b) is an averaged result for diffraction patterns of varying radius ($200$ to $350\,$nm in steps of $50\,$nm) with the same total scattered intensity. Due to the continuously varying local intensity ratio across the scattering angle, a different photon distribution is observed.}
    \label{fig:adu_dist_flat}
\end{figure*}

For each cluster with its associated ADU value $x_i$, we calculate the $S_{n,m}(x_i)$ using the pattern-specific one-peak heights and widths found by fitting Gaussians to the ADU histogram. The highest of these $S_{n,m}(x_i)$ is assumed to be the origin of the cluster. Summing all these different $S_{n,m}(x_i)$ gives a measure of the total value attributed to this cluster, from all considered combinations. A certainty value for the found combination is estimated by dividing the highest $S_{n,m}(x_i)$ by the total. The reconstructed components of all clusters with a certainty above a set threshold are then used to construct a red-only and a blue-only pattern.

\subsection*{Estimation of the probability of multiphoton events}

In the case of a recorded ADU value compatible with a signal of more than a photon, proper probability coefficients must be assigned to maximize the likelihood for a correct identification of the color's contribution. In this section, we discuss the case of two-photon events for the sake of simplicity, and the conclusions drawn here can be easily extended to higher photon counts. 

When a single photon is recorded, the probability $p$ of such a photon to be red is determined by the relative intensity of the red scattered field at the specific coordinates $i,j$ of the diffraction pattern, i.e.:
\begin{equation}\label{eq:redprob}
    p_{ij} = \frac{I^R_{ij}}{I^R_{ij} + I^B_{ij}} 
\end{equation}
where $I^R$ and $I^B$ are the intensities of the red and blue fields, respectively. The probability of recording a blue photon is, thus, $1-p_{ij}$.

The value of $p_{ij}$ fully determines the probabilities of multiple photon events. In particular, for a two-photon event, three probabilities have to be calculated: $p^{RR}_{ij}$ (two red photons), $p^{BB}_{ij}$ (two blue photons), and $p^{M}_{ij}$ (a mixed event). Their value is determined by a binomial distribution for two trials of independent events, that is:
\begin{equation}\label{eq:multiprob}
\begin{split}
p^{RR}_{ij} &= p^{RR} (p_{ij}) = p_{ij}^2 \\
p^{BB}_{ij} &= p^{BB} (p_{ij}) =  (1-p_{ij})^2 \\
p^{M}_{ij} &= p^{M} (p_{ij}) = 2\,p_{ij}\,(1-p_{ij}) 
\end{split}
\end{equation}

The probability of the two-photon events depends on the local probability $p_{ij}$, which requires the knowledge of the two intensity contributions $I^R_{ij}$ + $I^B_{ij}$, as shown in Eq.~\eqref{eq:redprob}. However, this information is inaccessible, and its retrieval is the primary focus of the separation method presented in this work. As a consequence, the coefficients in Eq.~\ref{eq:multiprob} have to be assigned \emph{a priori} and have to be agnostic of the local intensity of the fields.

The most straightforward option is to replace local probabilities with the average value:
\begin{equation}\label{eq:avgprob}
\hat{p}^X = \frac{1}{N^2} \sum_{i,j=1}^N  p^X_{ij} \sim \int_0^1 p^X (p) \, w(p) \, dp   
\end{equation}
where $N$ is the linear dimension of the diffraction pattern. The function $w(p)$ is a \emph{weight} function that expresses the relative abundance of pixels with a specific red probability $p$, i.e., a specific relative intensity of the red signal (through Eq. \eqref{eq:redprob}).

The integral form on the right-hand side of Eq.~\eqref{eq:avgprob} allows the approximation of $\hat{p}^{RR}$, $\hat{p}^{BB}$ and $\hat{p}^{M}$ through assumptions on ``how often'' the specific probability $p$ occurs in the diffraction data, i.e., on the weighting function $w(p)$.

For example, in the case of \emph{flat fields} (i.e., constant field intensity throughout the whole diffraction signal), only a single probability value $p_0$ is expected at all coordinates $ij$, $w(p) = \delta(p - p_0)$. As expected, when inserted into Eq.~\eqref{eq:avgprob}, it directly provides the canonical binomial coefficients $\hat{p}^{RR} = p_0^2 $, $\hat{p}^{BB} = (1-p_0)^2 $, and $\hat{p}^{M} = 2\,p_0\,(1-p_0)$. Specifically, for equal intensities of the two fields, the mixed photon event $\hat{p}^{M}$  is twice as probable as the single-color two-photon events $\hat{p}^{RR}$ and $\hat{p}^{BB}$. 
The simulation of such a homogeneous intensity distribution with a flat-field indeed yields the expected behavior, as illustrated in Supplementary Fig.~\ref{fig:adu_dist_flat}a. Here, the simulation is performed with equal intensities between the two colors. For the two-photon case, the mixed events have a count rate twice that of the single-color two-photon events.

However, the \emph{flat-field} assumption is significantly different from the behavior of the intensities expected in a diffraction pattern.
Due to the isolated nature of the samples, the diffraction intensities oscillate as a function of momentum transfer $\vec{q}$ between $0$ (known as \emph{nodes} of the scattered field) and a maximum value that decreases with increasing momentum transfer ($\propto \left|\vec{q} \right|^{-4}$ for hard spheres). Furthermore, the oscillation frequency of the signal on the detector is different between the two colors, due to the different wavelengths. For this reason, independent of the overall intensity ratio between the two colors, the \emph{weight} function $w(p)$ is non-zero in the whole interval $p \in [0,1]$. Specifically, there are always coordinates at which $I^B_{ij} \sim 0$ and $I^R_{ij} > 0$, corresponding to $p_{ij}=1$ as well as the inverse case, where $p_{ij}=0$. Due to the continuity of the scattering intensity, all other $p$ values are expected throughout the diffraction pattern.

In the case of similar overall intensities of the two colors, a model for $w(p)$ more accurate than the \emph{flat-field} discussed before can be obtained by assigning equal weights to all $p$ values, i.e., $w(p)=1$. In this case, the expectation values $\hat{p}^{RR}$, $\hat{p}^{BB}$, and $\hat{p}^{M}$ can be calculated through Eq.~\eqref{eq:avgprob} as:
\begin{equation}\label{eq:equalprob}
\hat{p}^{RR} = \hat{p}^{BB} = \hat{p}^{M} = \frac{1}{3}
\end{equation}

\begin{figure}[b]
    \centering
    \includegraphics[width=.9\linewidth]{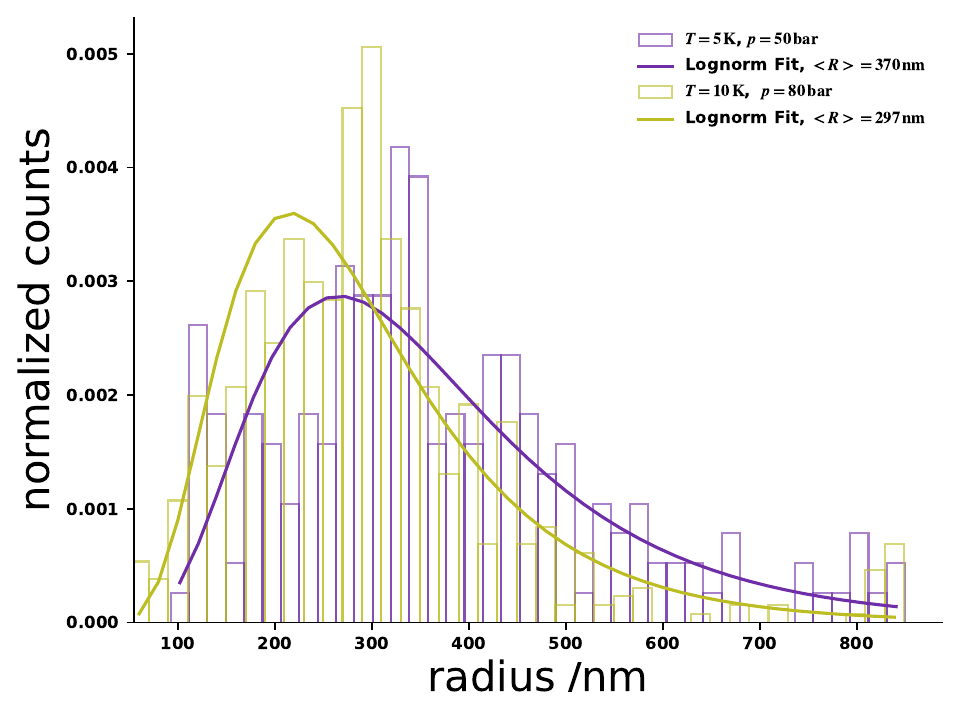}
    \caption{Size distribution of HNDs produced at two different valve temperatures $T$ and expansion pressures $p$, see the legend. For a pulsed nozzle, a lognormal distribution is assumed.}
    \phantomsection \label{fig:radius_dist}
\end{figure}

This conclusion can be further extended for cases where the overall intensity contribution of the two colors is not equal, i.e., $\sum_{ij} I^B_{ij} = \alpha \, \sum_{ij} I^R_{ij}$ for $\alpha \neq 1$. In these cases, the strictly non-zero nature of $w(p)$ throughout the interval $[0,1]$ is retained, with $0$ and $1$ still occurring at $I^R_{ij}=0$ and $I^B_{ij}=0$, respectively, independently of the scaling factor $\alpha$. Still, $\alpha<1$ increases the chances of $I^R_{ij}> I^B_{ij}$ locally (as the red signal is brighter), thus increasing $w(p)$ in the interval $[0,0.5]$ and decreasing $w(p)$ in $[0.5,1]$. The opposite behavior is expected in the case of $\alpha>1$. In those cases, the following relation approximately holds:
\begin{equation}\label{eq:imbprob}
\min[\hat{p}^{RR},\hat{p}^{BB}] \lesssim \hat{p}^{M} \lesssim \max[\hat{p}^{RR},\hat{p}^{BB}]
\end{equation}

These conclusions are reached with very generic assumptions about the shape of the scattered field, and in particular just the presence of zeros in the intensity contributions of the two colors and their similar decay as a function of the transfer momentum $|\vec{q}|$. The suitability of the condition deduced in Eq.~\eqref{eq:imbprob} is directly verified in Supplementary Fig.~\ref{fig:adu_dist_flat}b, which reports the histogram of the photon distributions up to three photon events for a simulated two-color diffraction pattern of a helium droplet with uneven color contributions.

The addition of more assumptions to achieve more restrictive conditions in $\hat{p}^{RR}$, $\hat{p}^{BB}$, and $\hat{p}^{M}$ would impact the generality of the pixel-based separation. 
For this reason, $\hat{p}^{RR}$ and $\hat{p}^{BB}$ are inferred directly from the ratio of single-photon events. The coefficient assigned to the mixed-color event is directly estimated as the average of $\hat{p}^{RR}$ and $\hat{p}^{BB}$, to fulfill the condition in Eq.~\eqref{eq:imbprob} without lacking generality. The same conclusions drawn here for two-photon events can be extended to higher photon counts, though these are less relevant to the overall performance of the separation algorithm.

\clearpage
\begin{sidewaysfigure}
\vspace{9cm}
\centering
    \begin{minipage}[c]{0.235\textwidth}
        \includegraphics[width=\textwidth]{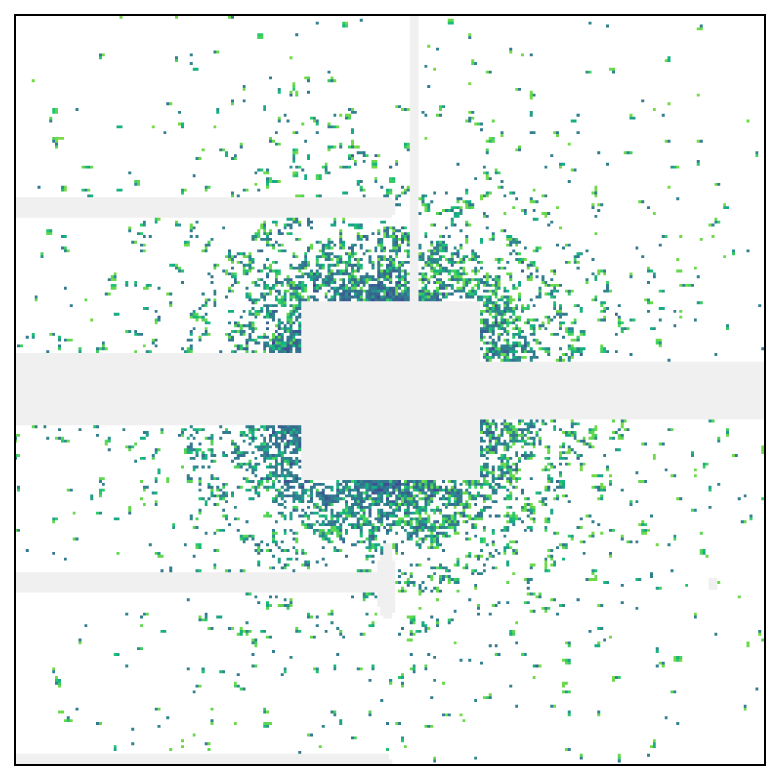}
        \phantomsection \label{fig:full1}
    \end{minipage}
    \hfill    
    \begin{minipage}[c]{0.235\textwidth}
        \includegraphics[width=\textwidth]{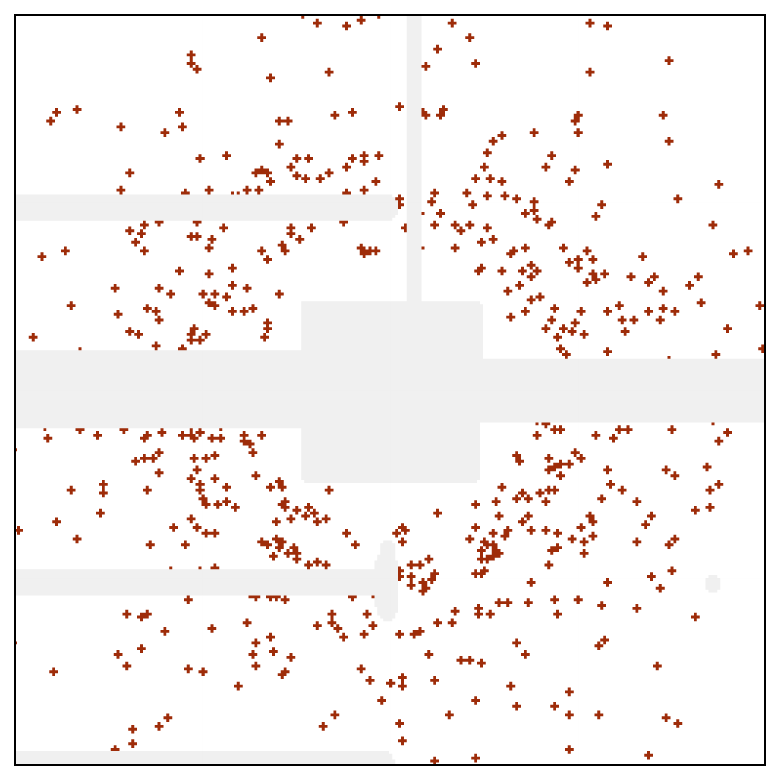}
        \phantomsection \label{fig:pump1}
    \end{minipage}
    \hfill    
    \begin{minipage}[c]{0.235\textwidth}
        \includegraphics[width=\textwidth]{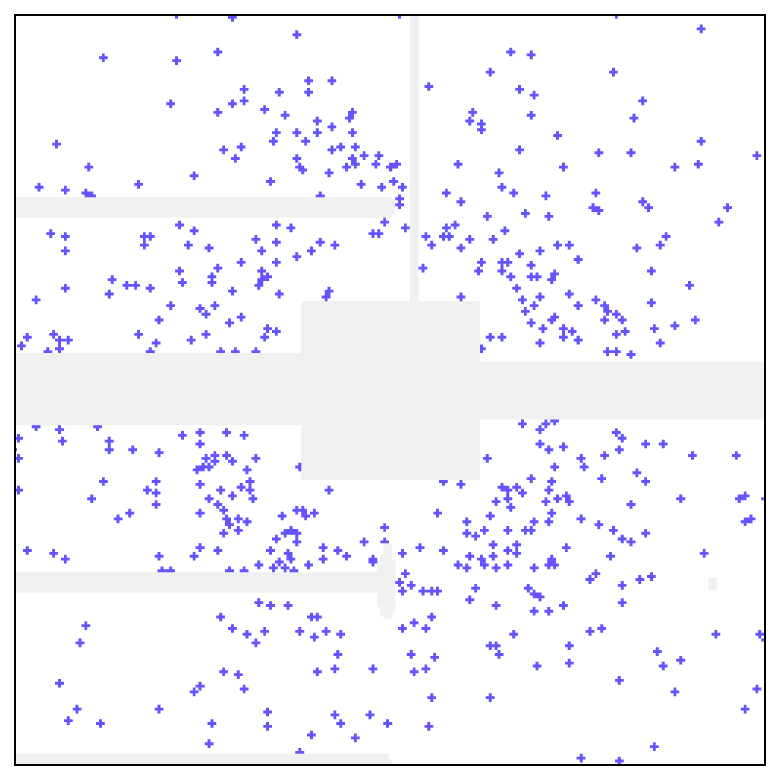}
        \phantomsection \label{fig:probe1}
    \end{minipage}
    \hfill
    \begin{minipage}[c]{0.26\textwidth}
        \vspace{.2cm}
        \includegraphics[width=\textwidth]{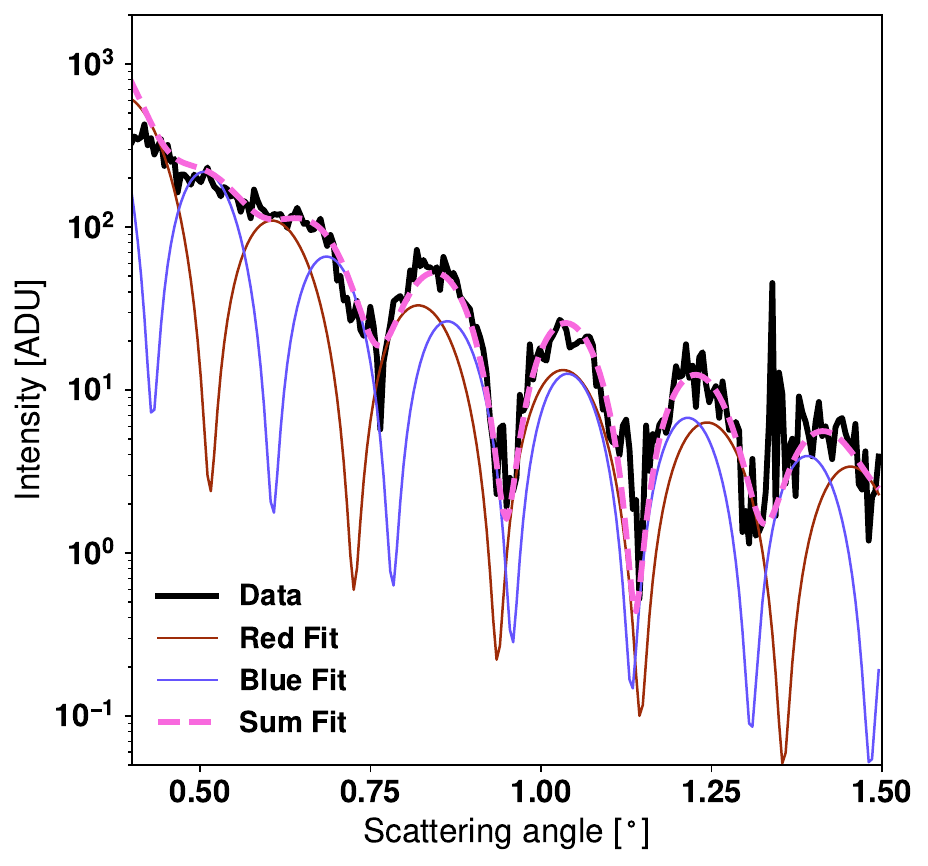}
        \phantomsection \label{fig:rad1}
    \end{minipage}
    \\
    \begin{minipage}[c]{0.235\textwidth}
        \includegraphics[width=\textwidth]{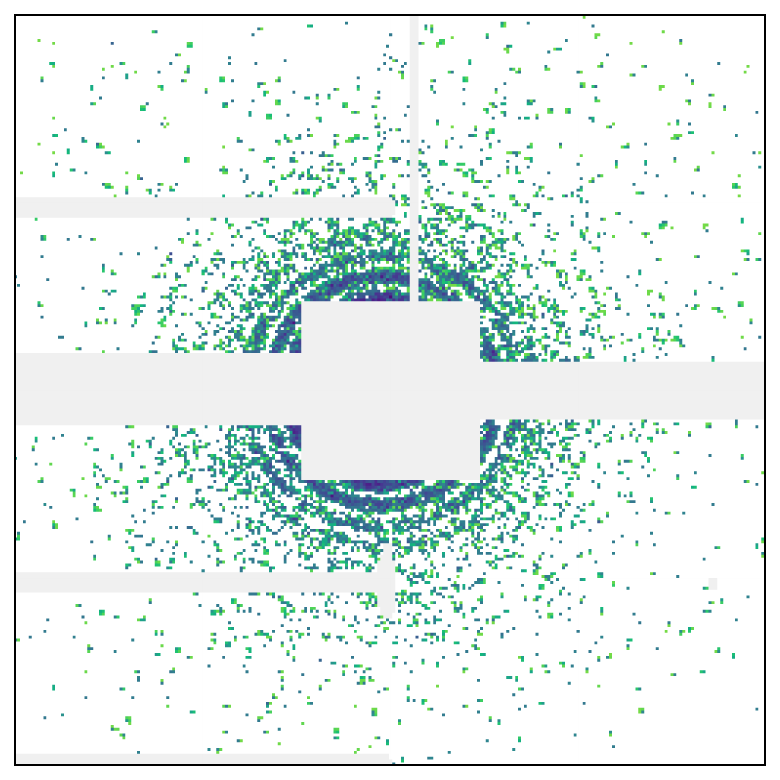}
        \phantomsection \label{fig:full2}
    \end{minipage}
    \hfill    
    \begin{minipage}[c]{0.235\textwidth}
        \includegraphics[width=\textwidth]{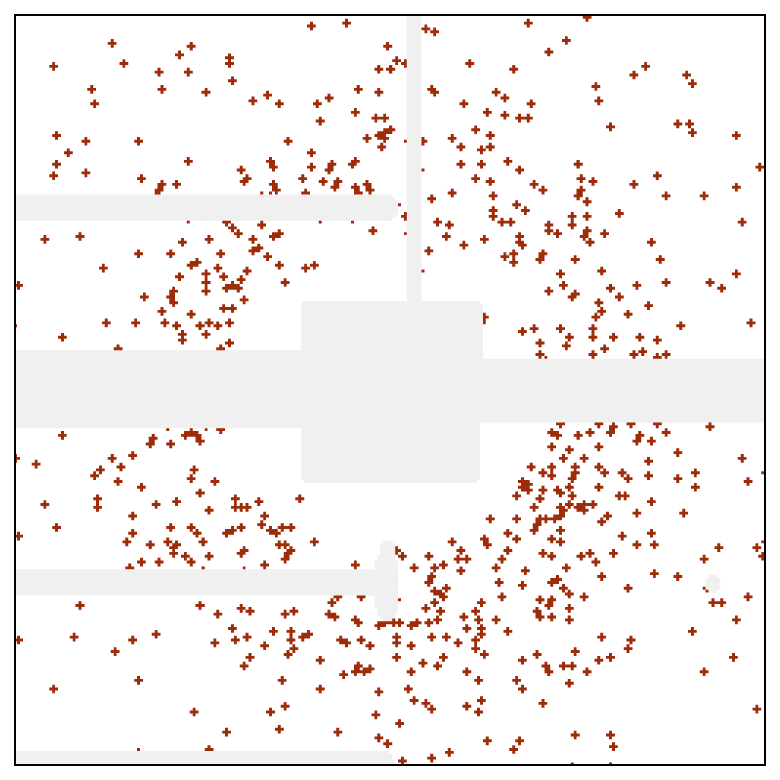}
        \phantomsection \label{fig:pump2}
    \end{minipage}
    \hfill    
    \begin{minipage}[c]{0.235\textwidth}
        \includegraphics[width=\textwidth]{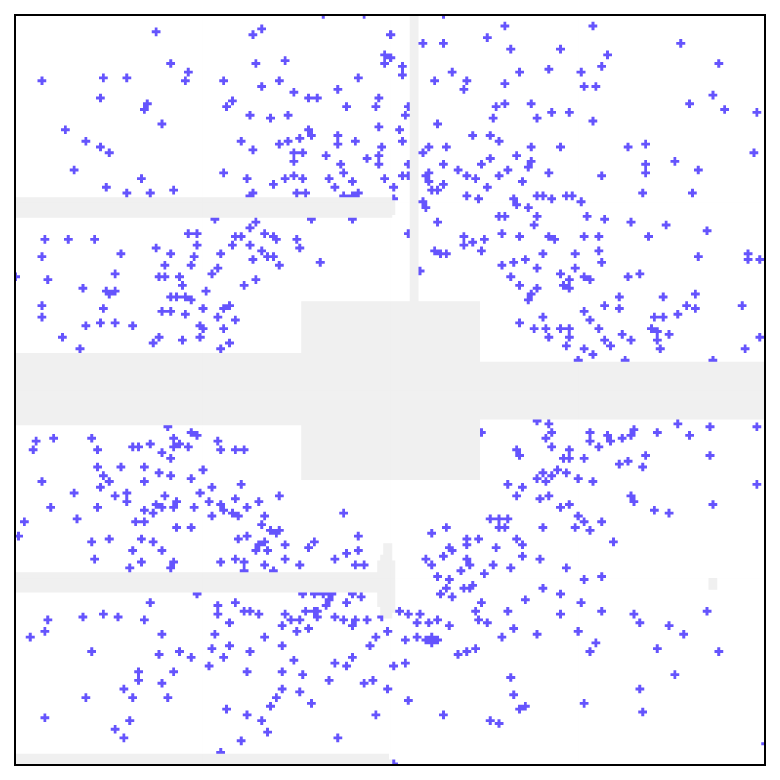}
        \phantomsection \label{fig:probe2}
    \end{minipage}
    \hfill
    \begin{minipage}[c]{0.26\textwidth}
        \vspace{.2cm}
        \includegraphics[width=\textwidth]{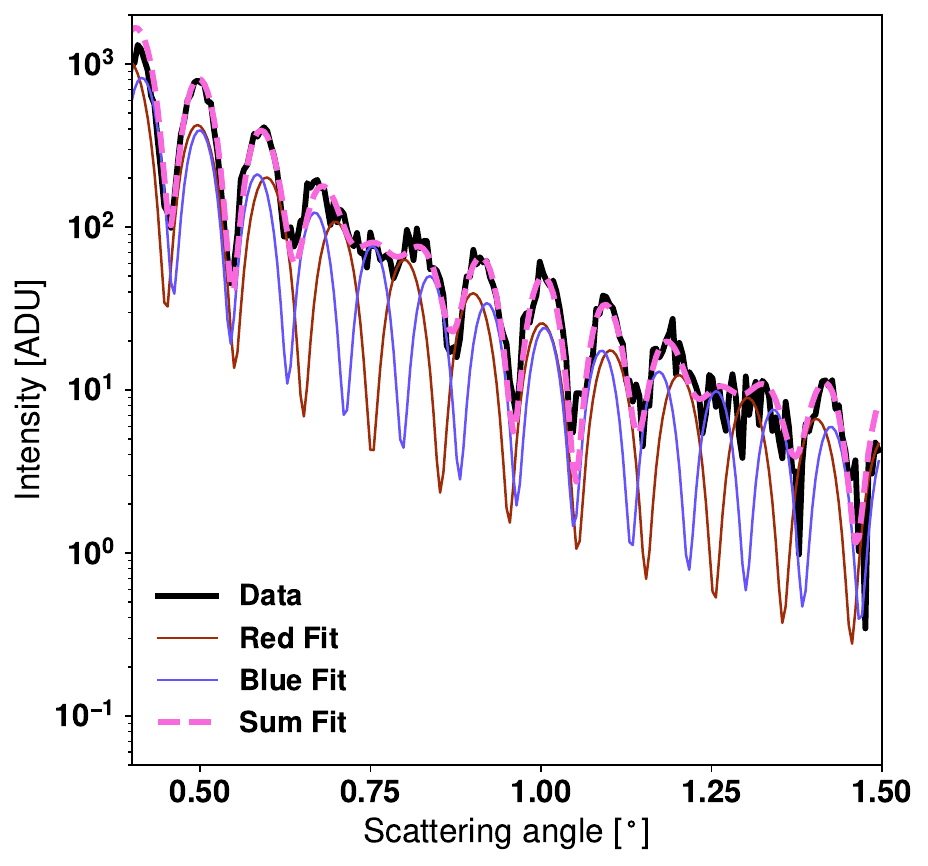}
        \phantomsection \label{fig:rad2}
    \end{minipage}
        \\
    \begin{minipage}[c]{0.235\textwidth}
        \includegraphics[width=\textwidth]{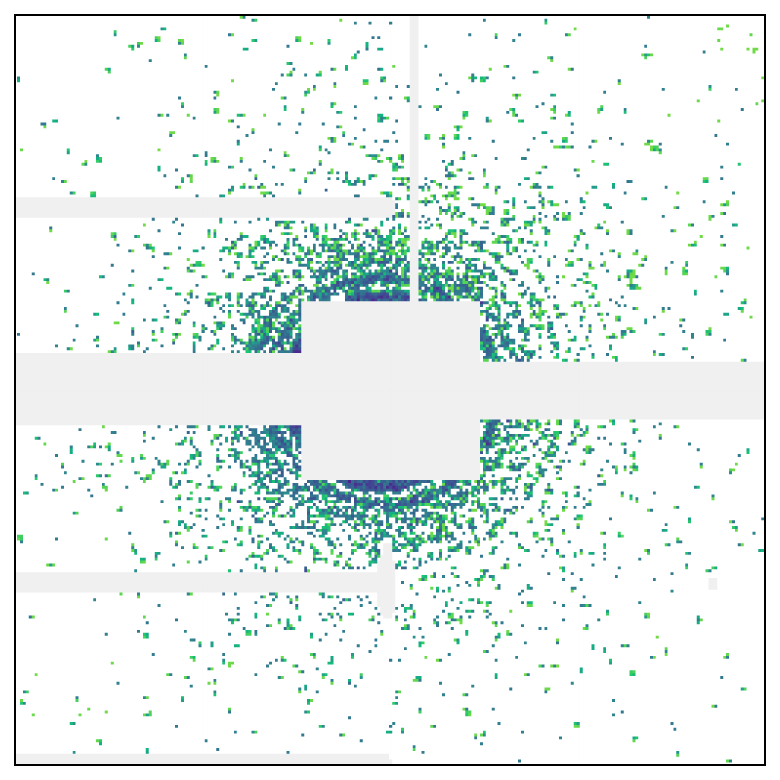}
        \phantomsection \label{fig:full3}
    \end{minipage}
    \hfill    
    \begin{minipage}[c]{0.235\textwidth}
        \includegraphics[width=\textwidth]{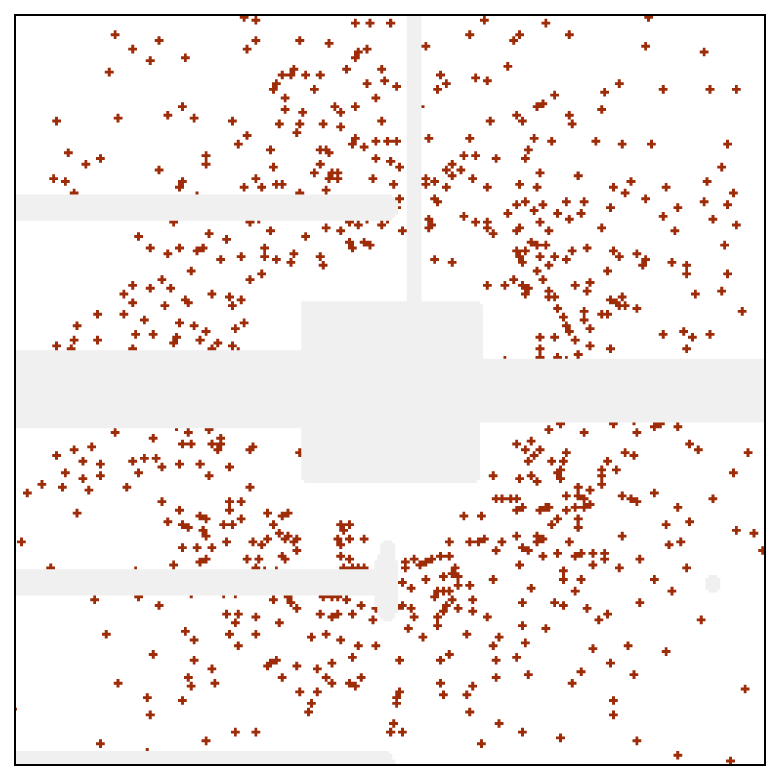}
        \phantomsection \label{fig:pump3}
    \end{minipage}
    \hfill    
    \begin{minipage}[c]{0.235\textwidth}
        \includegraphics[width=\textwidth]{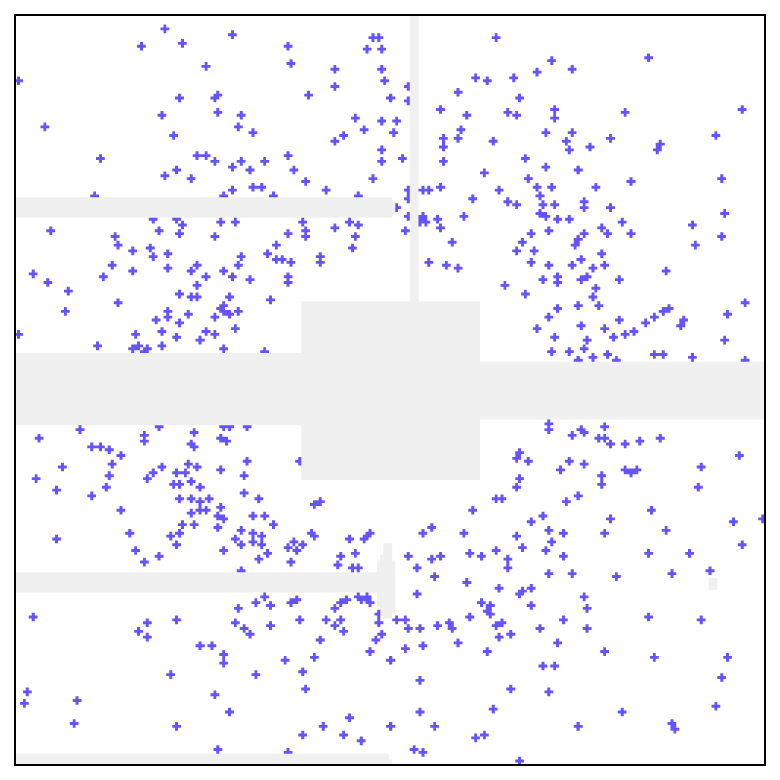}
        \phantomsection \label{fig:probe3}
    \end{minipage}
    \hfill
    \begin{minipage}[c]{0.26\textwidth}
        \vspace{.2cm}
        \includegraphics[width=\textwidth]{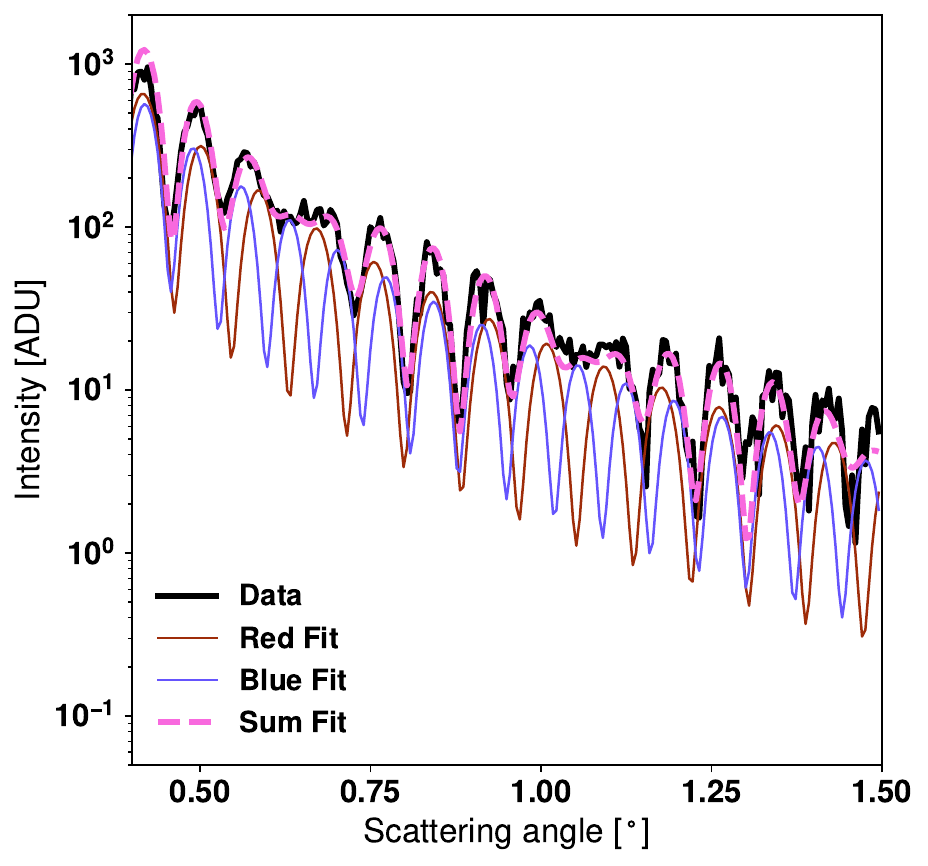}
        \phantomsection \label{fig:rad3}
    \end{minipage}
    \caption{Sample diffraction patterns and radial intensity profiles including Mie fits. Colormaps correspond to Figs. 1 and 2. The first column shows the full diffraction pattern. The second and third columns show the color-separated patterns. Here, we chose a low certainty threshold (1\%) to show the full separation capability. The last column depicts the radial intensity profiles; black lines are the original data, the pink dashed lines are the Mie fits, and the blue and red solid lines are the two color components resulting from the fit. Rows are ordered by HND size: $R = 173$, $354$, $423\, \unit{\nm}$.}
    \phantomsection \label{fig:exempImg}
\end{sidewaysfigure}
\clearpage
\begin{sidewaysfigure}
\vspace{9cm}
\centering
    \begin{minipage}[c]{0.235\textwidth}
        \includegraphics[width=\textwidth]{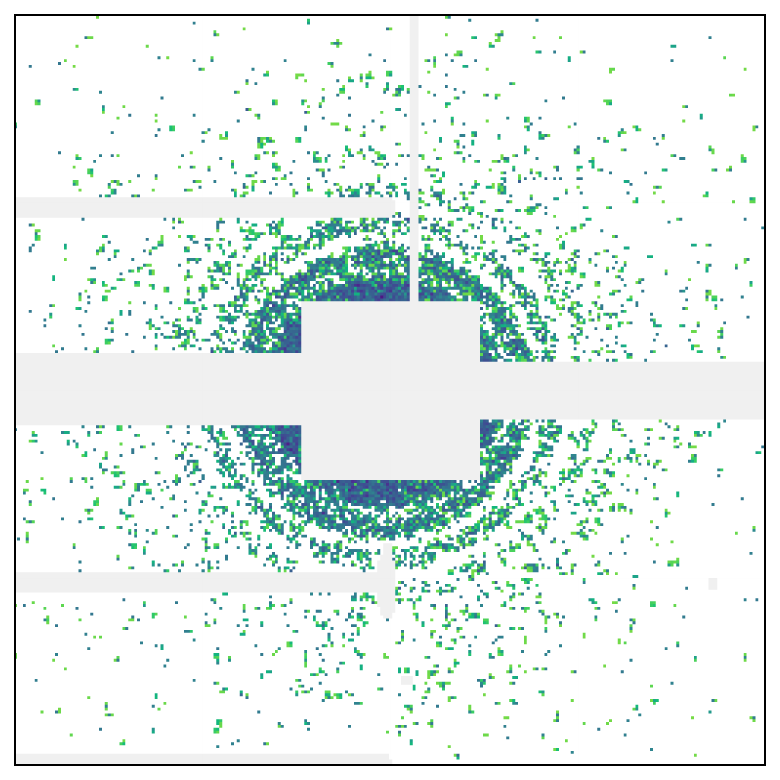}
        \phantomsection \label{fig:full4}
    \end{minipage}
    \hfill    
    \begin{minipage}[c]{0.235\textwidth}
        \includegraphics[width=\textwidth]{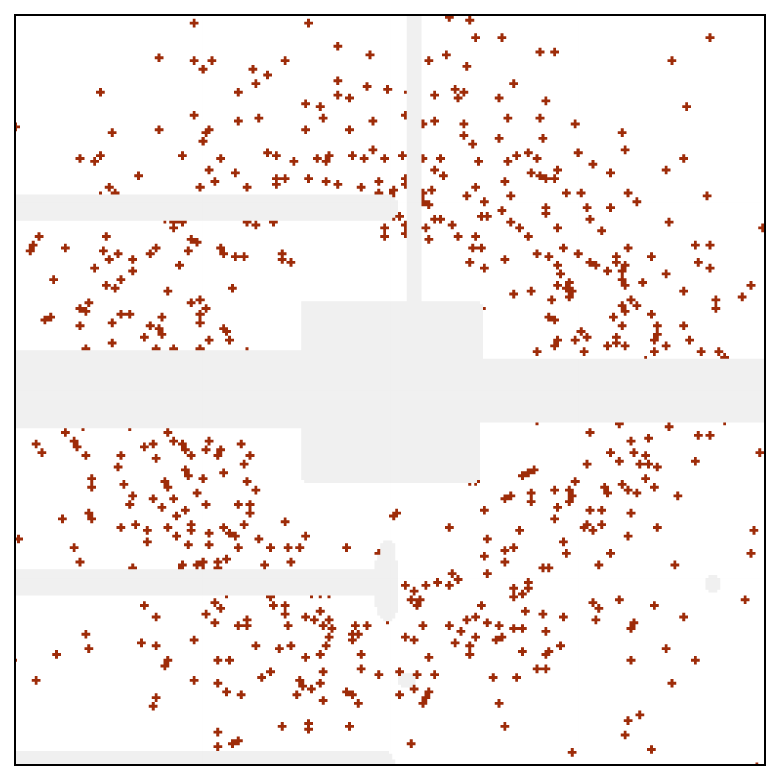}
        \phantomsection \label{fig:pump4}
    \end{minipage}
    \hfill    
    \begin{minipage}[c]{0.235\textwidth}
        \includegraphics[width=\textwidth]{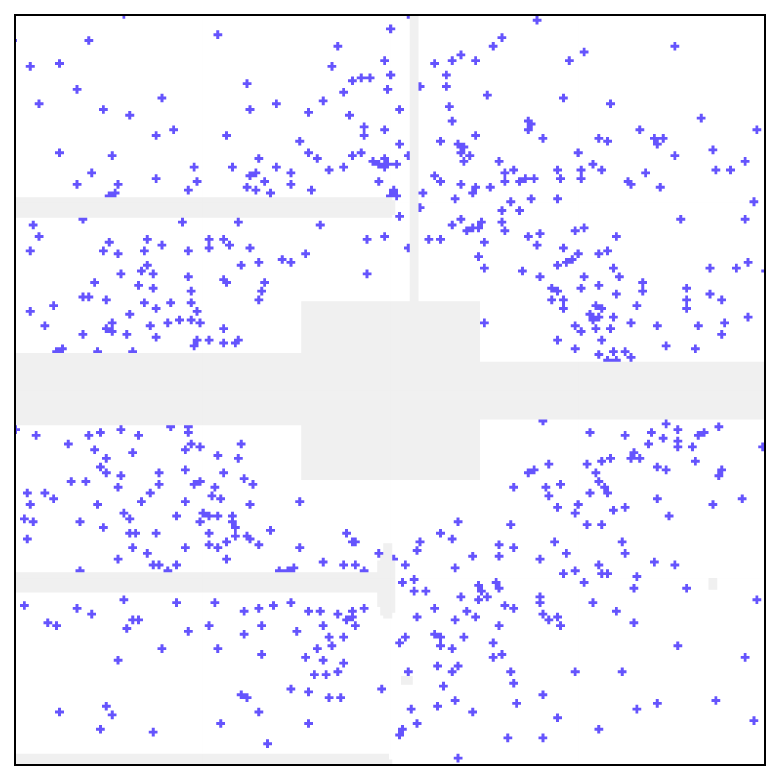}
        \phantomsection \label{fig:probe4}
    \end{minipage}
    \hfill
    \begin{minipage}[c]{0.26\textwidth}
        \vspace{.2cm}
        \includegraphics[width=\textwidth]{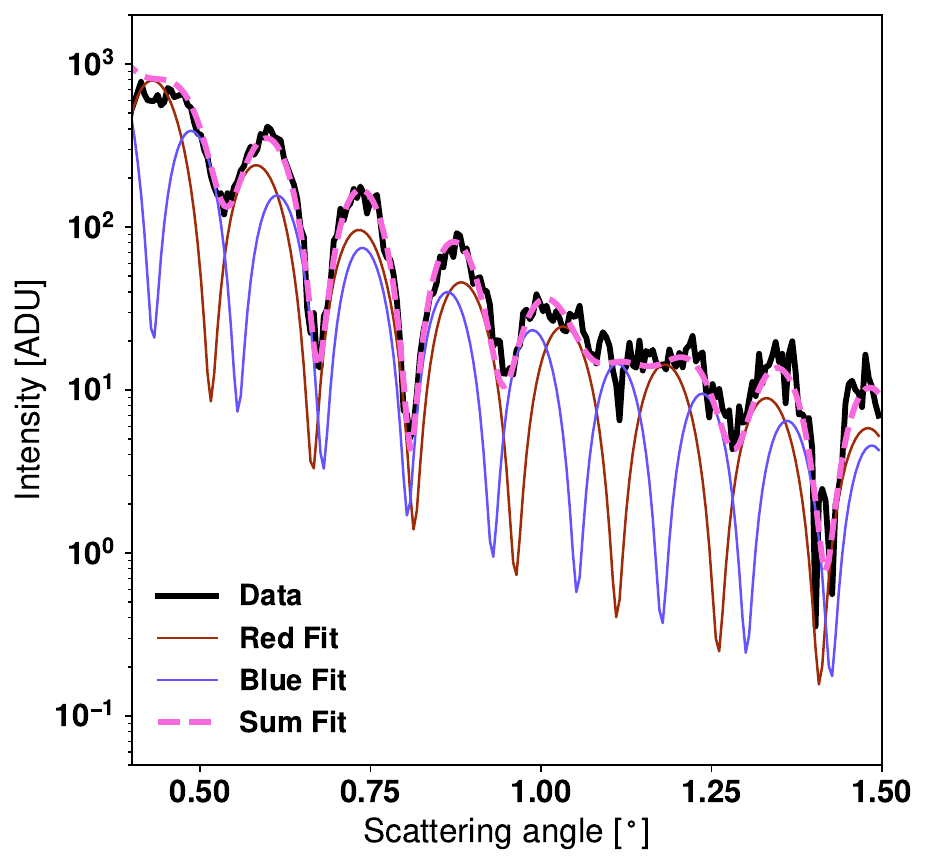}
        \phantomsection \label{fig:rad4}
    \end{minipage}
    \\
    \begin{minipage}[c]{0.235\textwidth}
        \includegraphics[width=\textwidth]{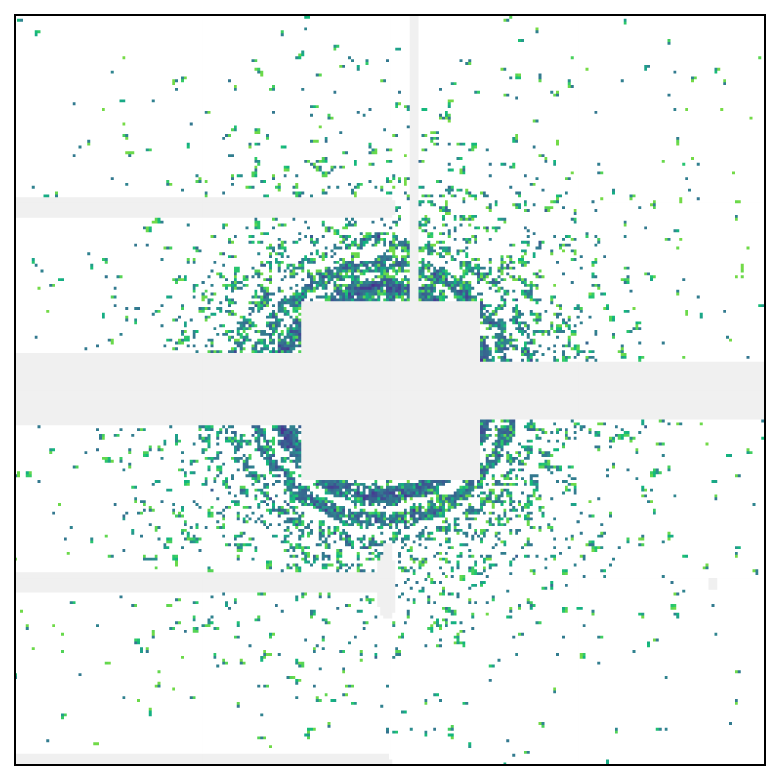}
        \phantomsection \label{fig:full5}
    \end{minipage}
    \hfill    
    \begin{minipage}[c]{0.235\textwidth}
        \includegraphics[width=\textwidth]{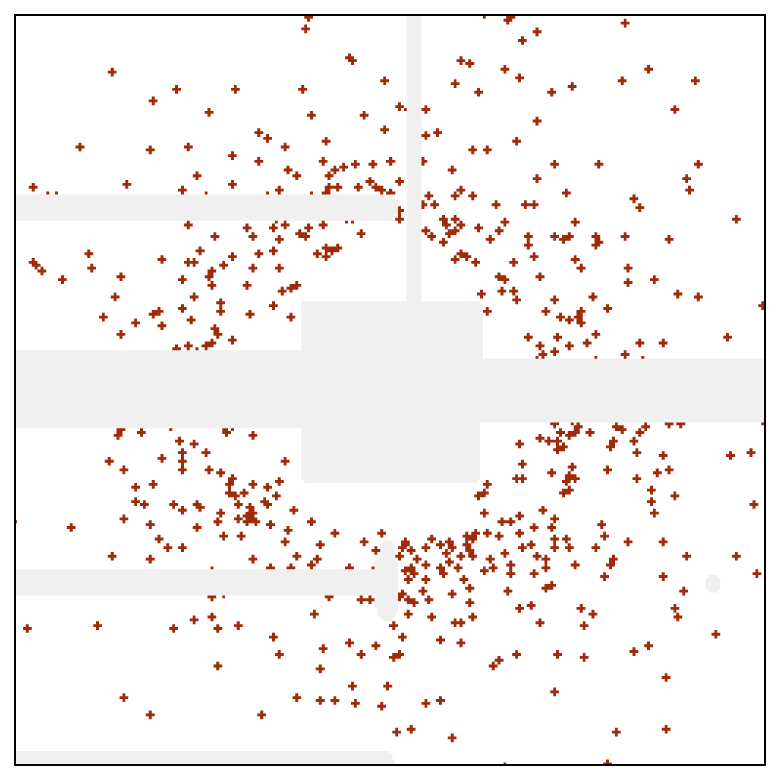}
        \phantomsection \label{fig:pump5}
    \end{minipage}
    \hfill    
    \begin{minipage}[c]{0.235\textwidth}
        \includegraphics[width=\textwidth]{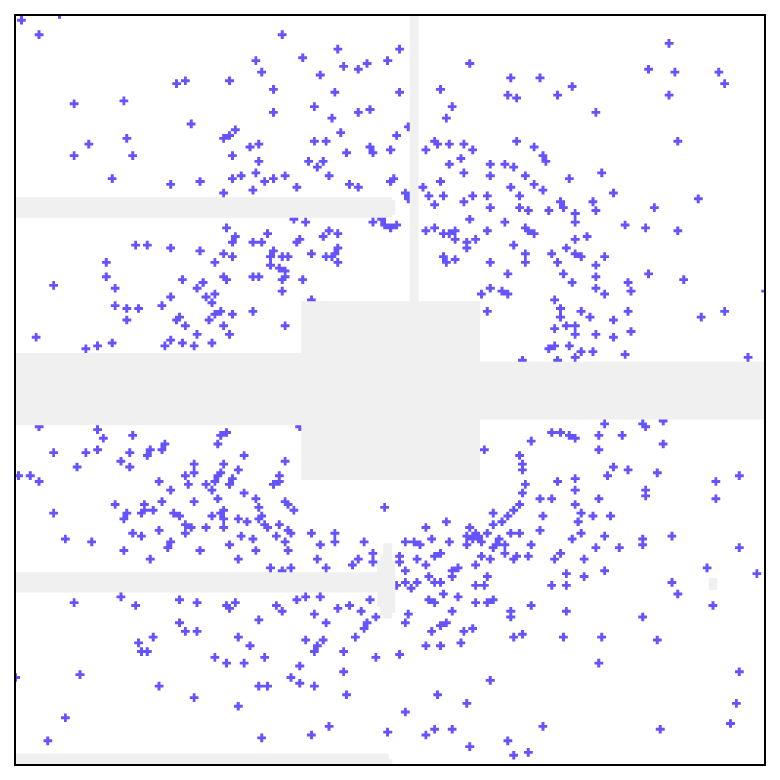}
        \phantomsection \label{fig:probe5}
    \end{minipage}
    \hfill
    \begin{minipage}[c]{0.26\textwidth}
        \vspace{.2cm}
        \includegraphics[width=\textwidth]{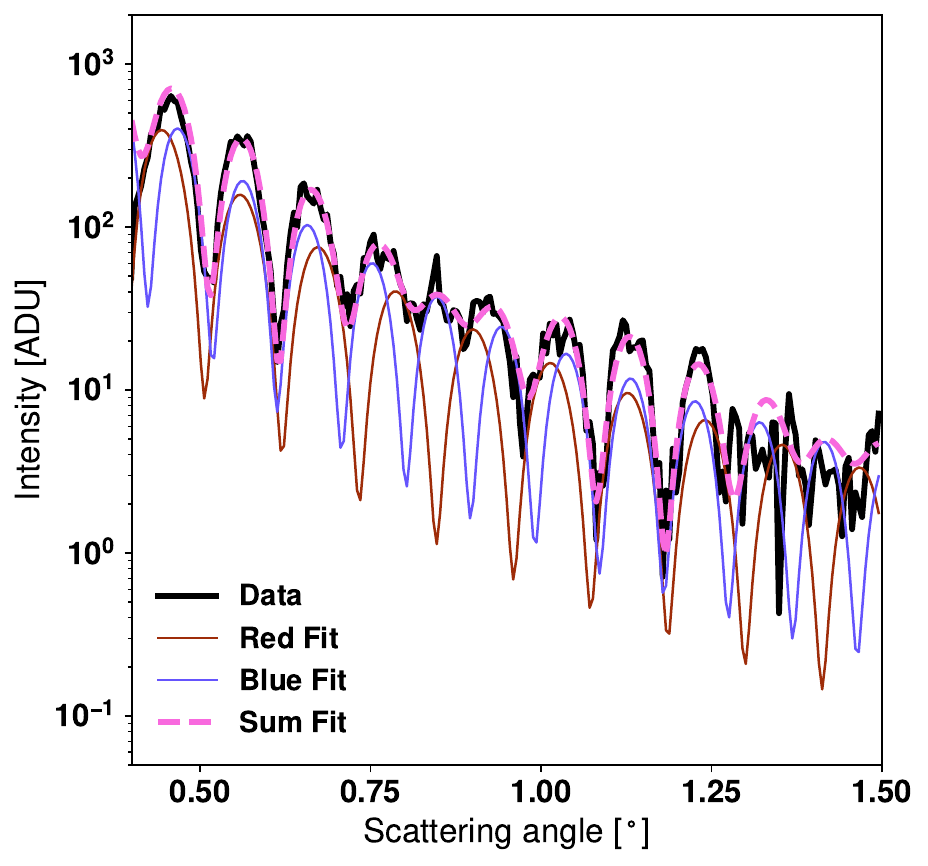}
        \phantomsection \label{fig:rad5}
    \end{minipage}
        \\
    \begin{minipage}[c]{0.235\textwidth}
        \includegraphics[width=\textwidth]{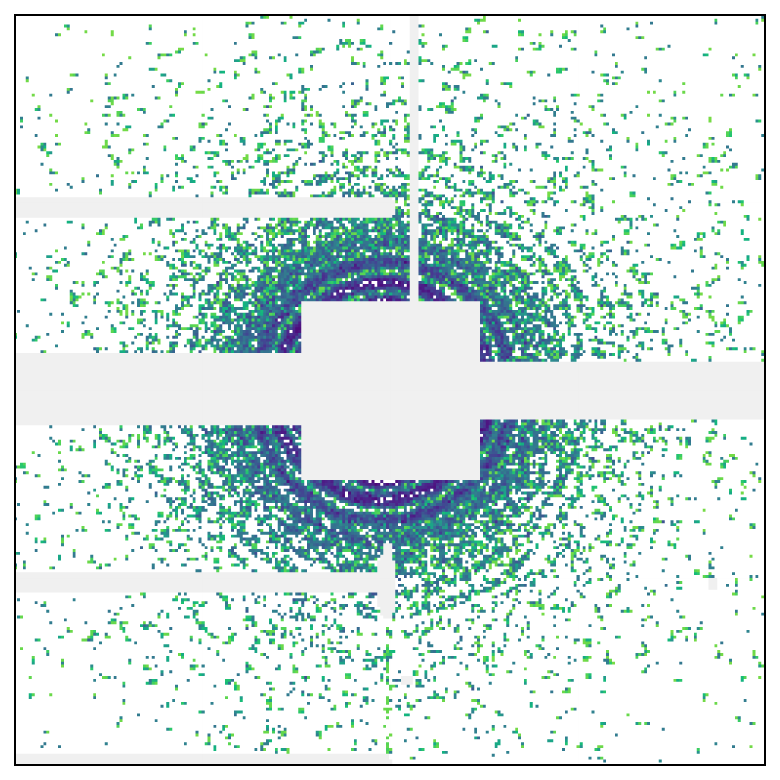}
        \phantomsection \label{fig:full6}
    \end{minipage}
    \hfill    
    \begin{minipage}[c]{0.235\textwidth}
        \includegraphics[width=\textwidth]{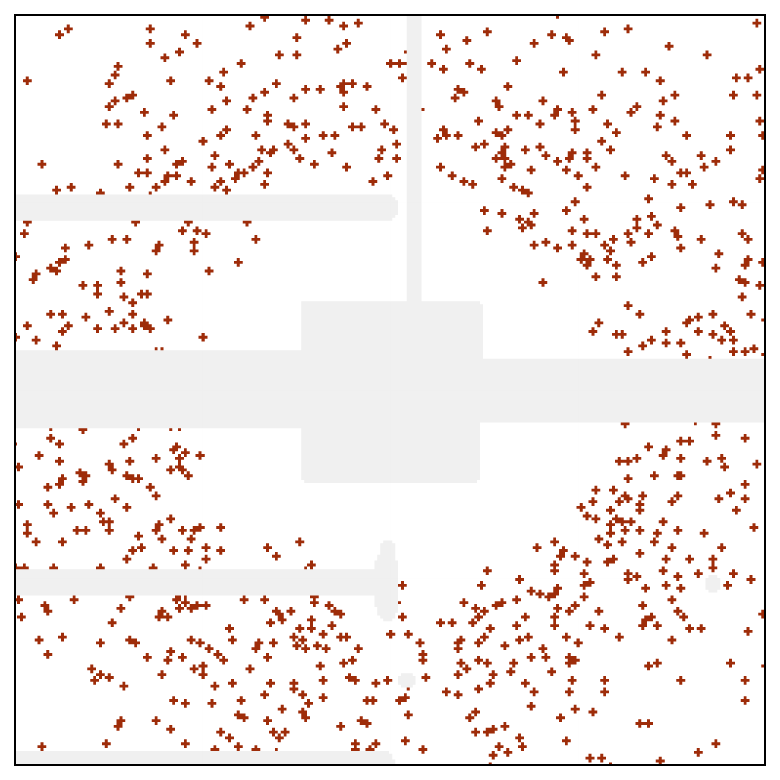}
        \phantomsection \label{fig:pump6}
    \end{minipage}
    \hfill    
    \begin{minipage}[c]{0.235\textwidth}
        \includegraphics[width=\textwidth]{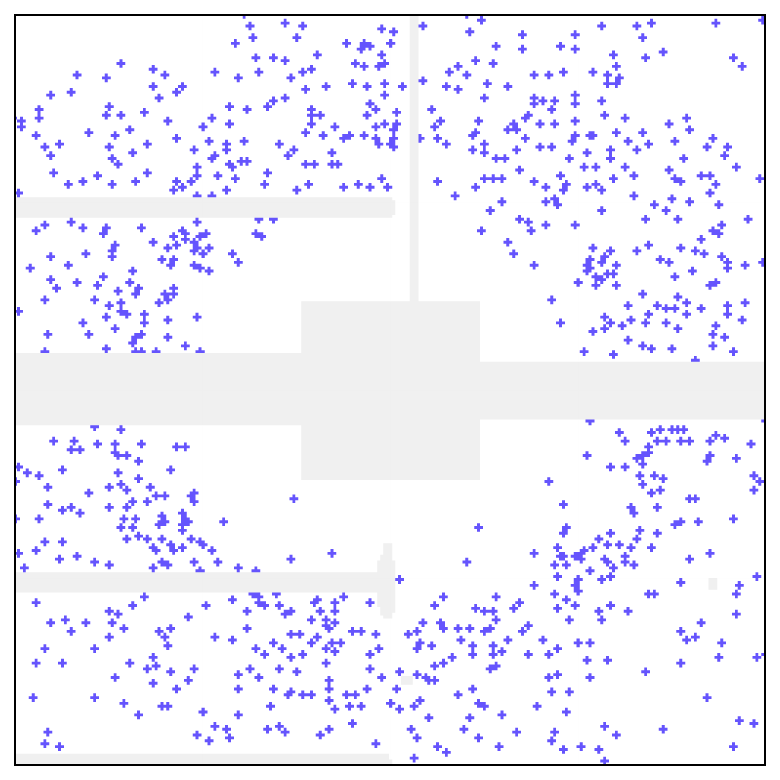}
        \phantomsection \label{fig:probe6}
    \end{minipage}
    \hfill
    \begin{minipage}[c]{0.26\textwidth}
        \vspace{.2cm}
        \includegraphics[width=\textwidth]{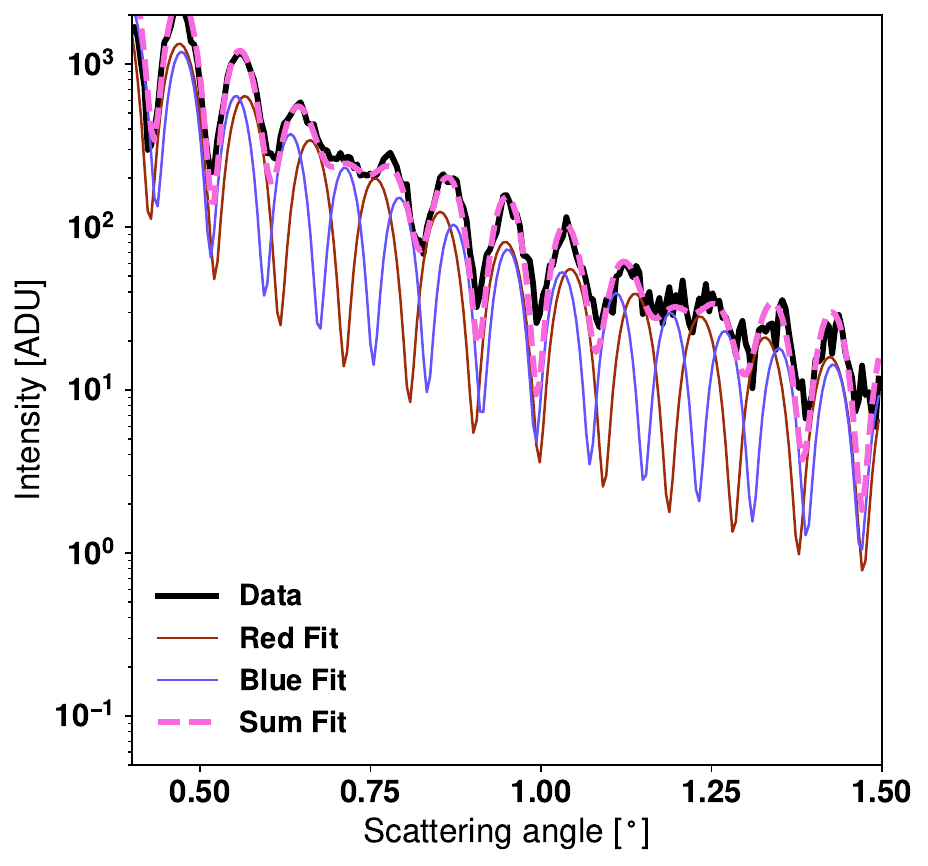}
        \phantomsection \label{fig:rad6}
    \end{minipage}
    \caption{Sample diffraction patterns and radial intensity profiles including Mie fits: Figure is structured as Supplementary Fig.~\ref{fig:exempImg}.\\ Rows are ordered by HND size: $R = 239$, $314$, $374\, \unit{\nm}$.}
    \phantomsection \label{fig:exempImg2}
\end{sidewaysfigure}
\clearpage

\makeatletter
\begin{@twocolumnfalse}
  \centering

  \includegraphics[width=0.90\textwidth]{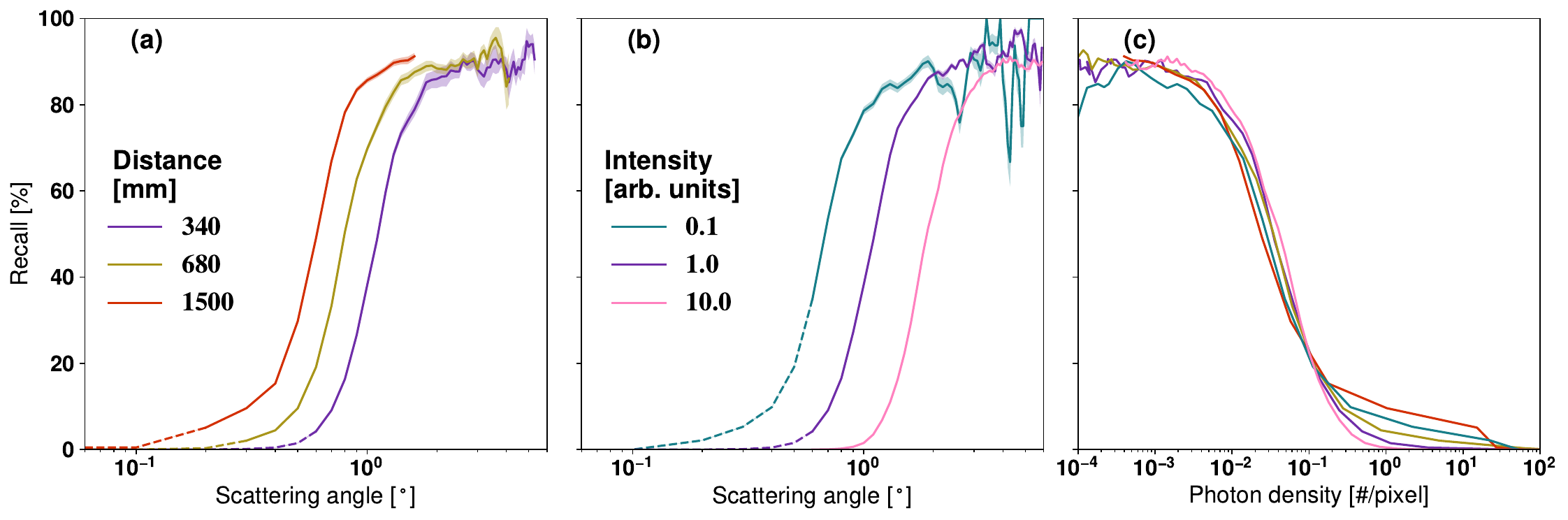}

  \refstepcounter{figure}
  \label{fig:Recall}

  \vspace{0.5em}
  \begin{minipage}{0.92\textwidth}
    \small
    \noindent\textbf{Supplementary Fig.~\thefigure.}
    (a) Recall (efficiency of correctly assigning pixel values to noise or to single or multiple red and blue photons) as a function of the scattering angle. Three different positions of the detector with respect to the interaction point are assumed. The experiment's distance was $340$~mm. (b) Recall for various intensities of the pulses relative to the experimental value set to $1$. Dashed lines indicate the ranges of angles that are excluded by the gap between detector modules in the experiment. (c) Same data as in (a) and (b) replotted as a function of the area density of photons on the detector.
  \end{minipage}

  \vspace{1.0em}
\end{@twocolumnfalse}
\makeatother

\subsection*{Influences on the recall}

The recall is not constant throughout the entire diffraction pattern but rather depends on the local signal intensity. Supplementary Fig.~\ref{fig:Recall} shows the recall as a function of the scattering angle for different values of the distance between the detector and the interaction point (a) and for different values of the intensity of the pulses relative to the experimental value (b). The experimental conditions are specified with a distance of $340$~mm and an intensity value of $1$.

The area of high recall could in principle be extended by increasing the distance to the detector, Supplementary Fig.~\ref{fig:Recall}a, and by reducing the intensity of scattered light (b). However, the latter goes at the expense of the signal-to-noise ratio, as seen from the large scatter of the data for a relative intensity of $0.1$ (red line). This translates to a lower spatial resolution as the resolution is limited by the maximum scattering angle at which diffraction data are recorded with sufficient statistics. Similarly, by moving the detector further back, not only are the scattered photons dispersed over more pixels, which increases the recall, but also the large-angle scattering signal becomes too sparse, and the maximum usable scattering angle is reduced.

When we replot all the data shown in (a) and (b) as a function of the area density of scattered photons on the detector, i.e., the mean number of photons per pixel, all curves overlap almost perfectly, see Supplementary Fig.~\ref{fig:Recall}c. Thus, the photon density is the decisive quantity that should be kept below $0.1$ as the recall steeply drops to zero above this value.

\begin{figure}
\vspace{7.5cm}
    \centering
        \begin{minipage}{.3\textwidth}
        \makebox[\linewidth]{
            \includegraphics[height=0.22\textheight]{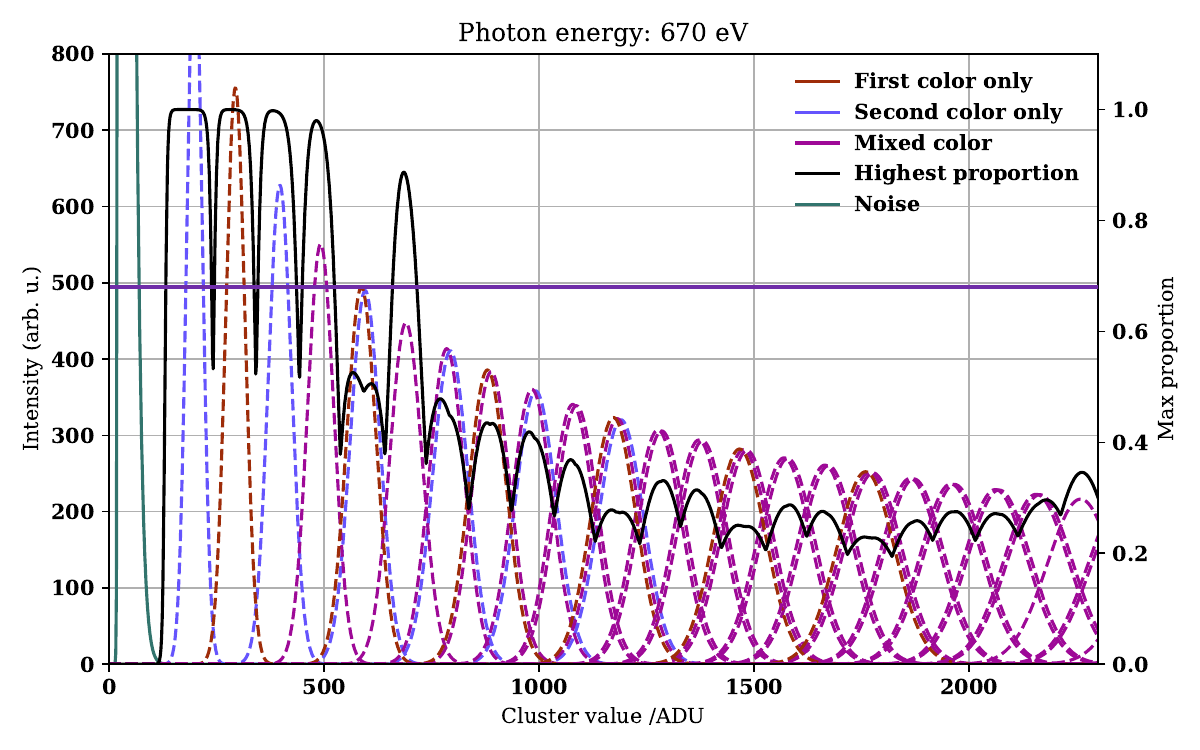}}
        \end{minipage}
        \begin{minipage}{.3\textwidth}
        \makebox[\linewidth]{
            \includegraphics[height=0.22\textheight]{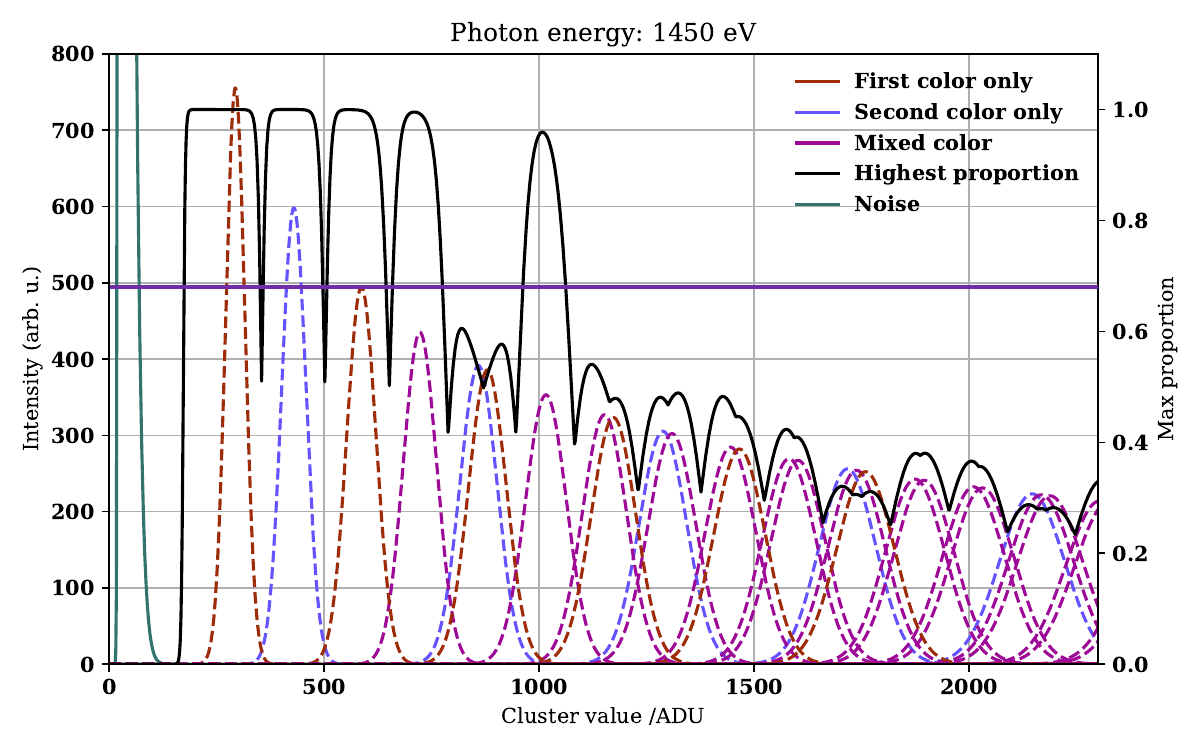}}
        \end{minipage}
        \begin{minipage}{.3\textwidth}
        \makebox[\linewidth]{
            \includegraphics[height=0.22\textheight]{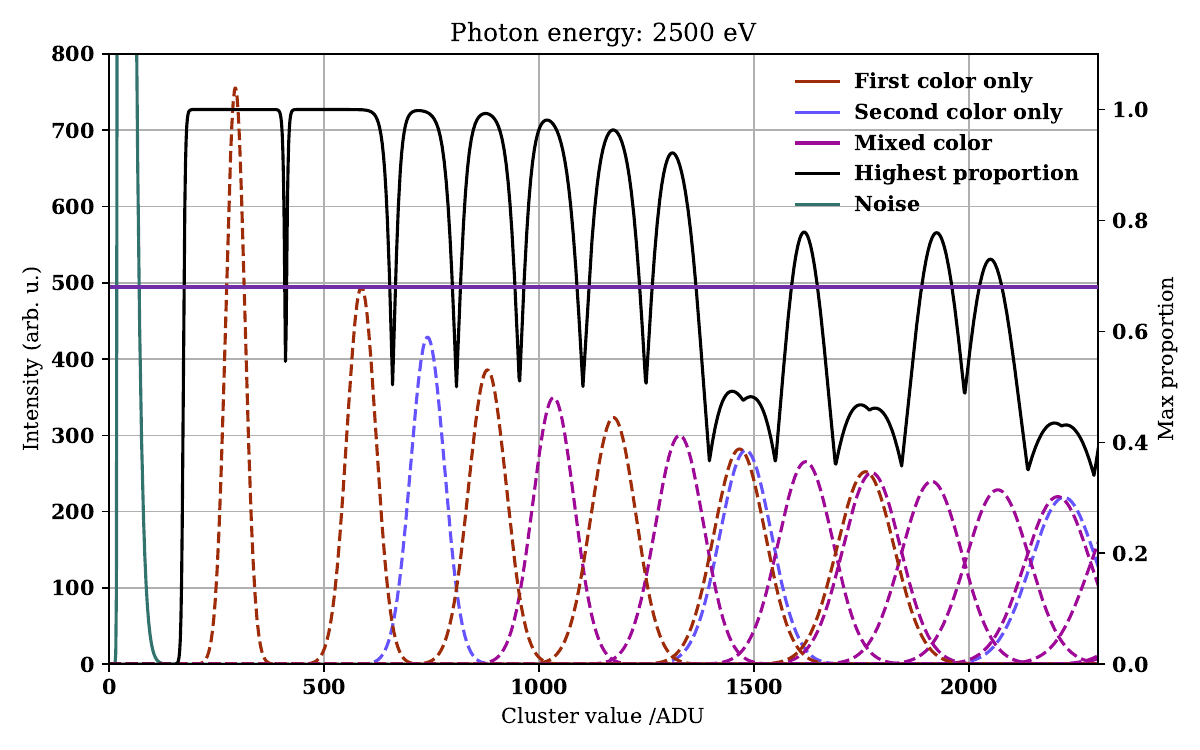}}
        \end{minipage}
    \caption{Simulated distributions of pixel values as shown and explained in Fig.~8 (in the Methods section) for the combination of photon energies $E_\mathrm{ph,\,1}=1000$~eV and $E_\mathrm{ph,\,2}=$~670, 1450, and 2500~eV.}
    \phantomsection \label{fig:difPhoEne}
\end{figure}
\begin{figure}
    \centering
    \begin{minipage}{0.47\textwidth}
    \makebox[\linewidth]{
        \includegraphics[width=\linewidth]{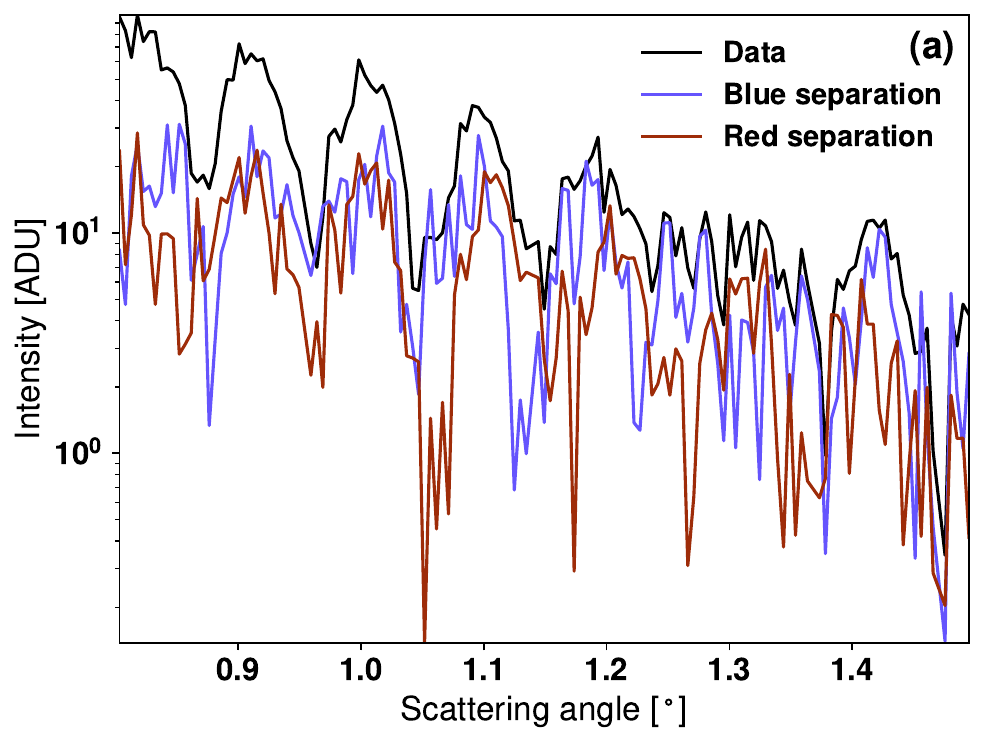}}
    \end{minipage}
    \begin{minipage}{0.47\textwidth}
    \makebox[\linewidth]{
        \includegraphics[width=\linewidth]{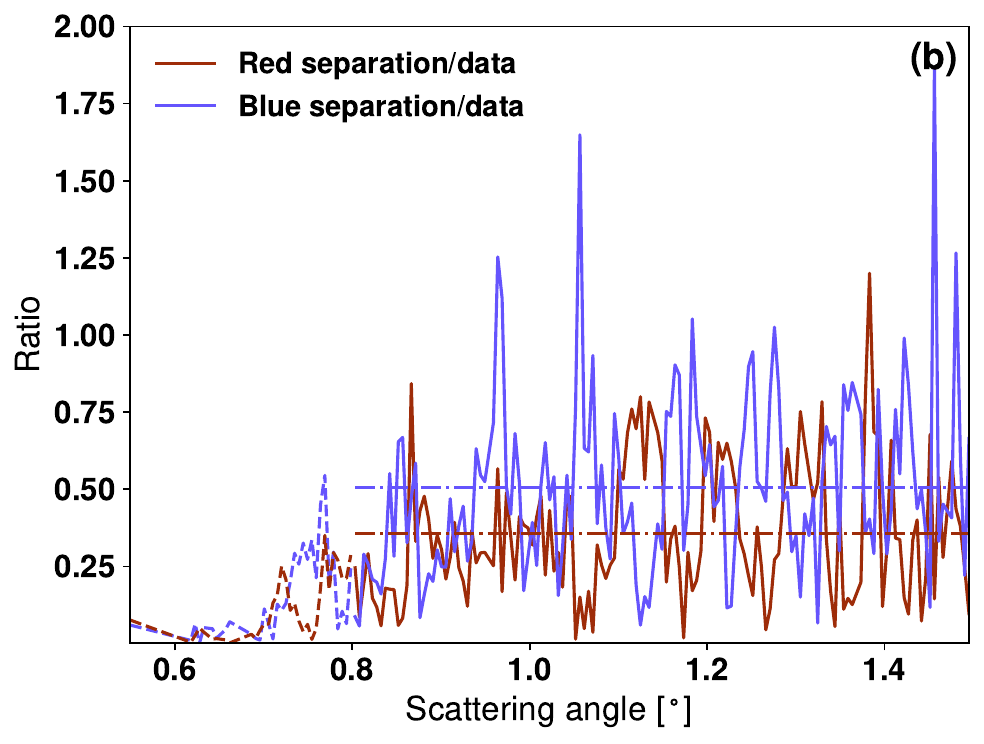}}
    \end{minipage}
    \caption{(a) Radial intensity profiles of a selected diffraction pattern where the fluence of red and blue X-ray pulses was approximately equal. The black line is the profile of the total two-color pattern, the red and blue lines are the profiles of the pixel-based color-separated patterns. (b) Ratio of radial intensities of color-separated patterns with respect to the full pattern. At scattering angles $<0.8\,^\circ$ (dashed part), the color separation is inefficient, whereas at angles $>0.8\,^\circ$, average relative intensities of 0.36/0.51 for the red/blue pulse are reached, marked with the horizontal lines.
    }
    \phantomsection \label{fig:relInt}
\end{figure}

\begin{figure}
    \centering
    \includegraphics[width=1\linewidth]{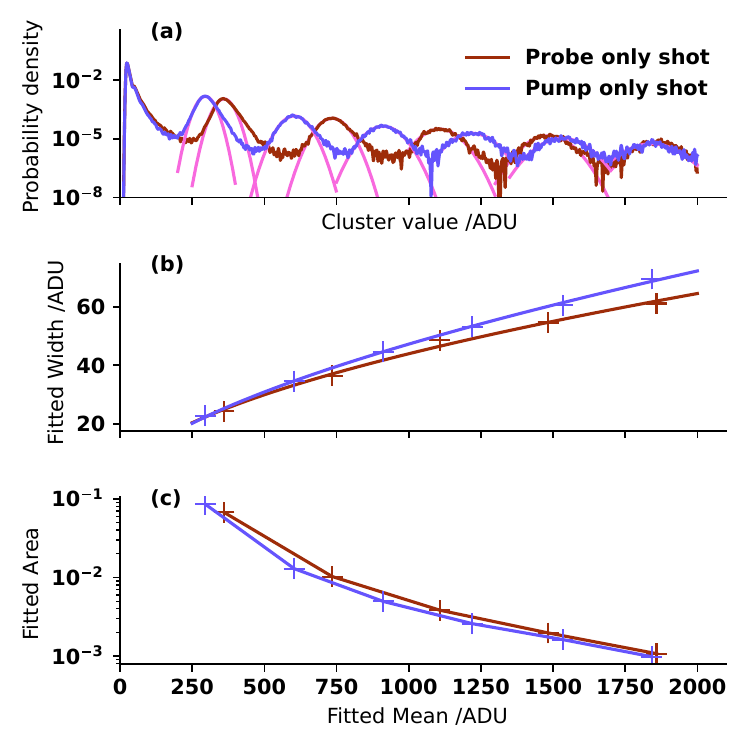}
    \caption{Analysis of the distribution of pixel counts for a set of measured patterns recorded using ``blue pulses'' only. The areas of the one-photon, two-photon etc. peaks in the histograms in (a) quickly drop as a function of the number of photons, (c). The widths of the peaks increase with the intensity as $\bar{I}^{0.55}$ for the red and as $\bar{I}^{0.6}$ for the blue pulses, (b).}
    \phantomsection \label{fig:ADU_width}
\end{figure}

\begin{figure}
    \centering
    \includegraphics[width=1\linewidth]{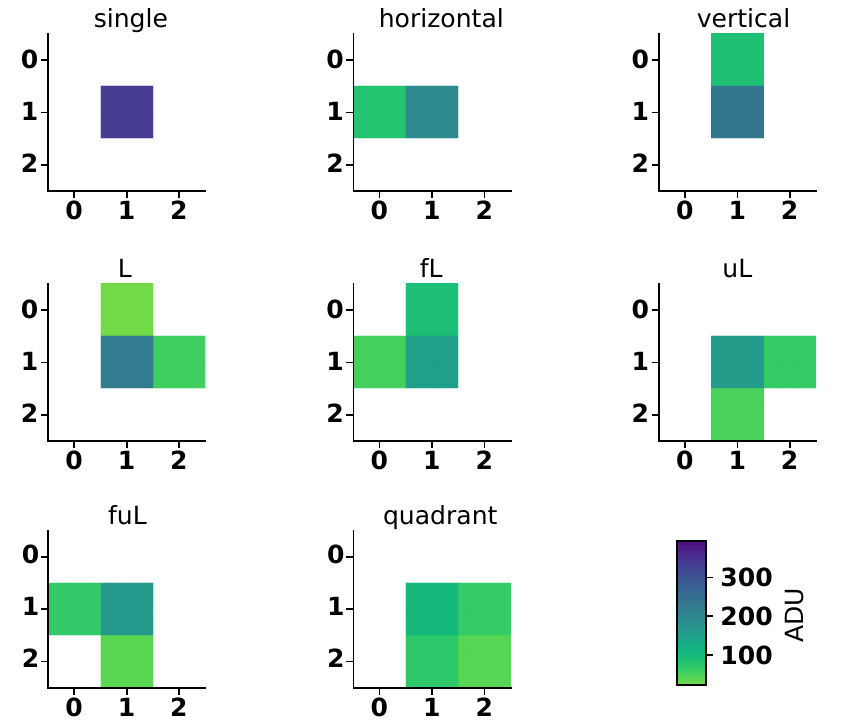}
    \caption{All possible pixel patterns produced by a single photon.}
    \phantomsection \label{fig:shapes}
\end{figure}

\begin{table}[]
\caption{Overview of single photon patterns found in the brightest 75 shots. Naming as in Supplementary Fig.~\ref{fig:shapes}.}
\phantomsection \label{tab:pat}
\centering
\begin{tabular}{r|l}
Shape & Fraction [\%]\\
\hline
single & 46 \\
horizontal & 21 \\
vertical & 17 \\
fuL & 3 \\
uL & 3 \\
L & 3 \\
fL & 3 \\
quadrant & 4 \\
\end{tabular}
\end{table}
\FloatBarrier

\clearpage

\clearpage
\onecolumngrid

\begingroup
\setlength{\parindent}{1em}

\begin{minipage}[t]{0.48\textwidth}
\setlength{\parindent}{1em}
\subsection*{Bounding the positional shift in cluster assignment}
\vspace{-.3cm}
\indent Assigning a cluster places it at its center-of-mass pixel. While this introduces a slight coordinate shift in the diffraction pattern, this effect is limited since the procedure is only feasible for small, few-photon clusters. To determine an upper bound for this shift, we calculated the Euclidean distance from the center-of-mass pixel to the furthest extent of the original pattern. These are shown in Supplementary Fig.~\ref{fig:upper_bound_for_shift} for 10 randomly chosen diffraction patterns processed with 50\% confidence. As shown, for the vast majority of patterns, the actual photon impact point is shifted by at most $\sqrt{2}$ pixels from the assigned pixel coordinate. This results in an average shift of $0.8$ pixels using this conservative upper bound estimate.
\vspace{-.5cm}
\subsection*{Impact of positional uncertainty on structural resolution}
\vspace{-.4cm}
Our assignment procedure introduces a localized shift of approximately $0.8$ pixels. In this section, we briefly discuss how this uncertainty in photon position does not affect the information content of the diffraction data. The discussion involves three aspects: signal oversampling, geometric error scaling, and physical bandwidth limits.

In single-particle CDI, the maximum frequency of signal variation on the detector is strictly constrained by the sample's spatial extent. The Fourier transform of the diffraction pattern corresponds to the sample's autocorrelation function. For an isolated sample of radius $R$, the autocorrelation has a maximum extension of $D_\mathrm{max}=4R$. Consequently, the fastest oscillation (fringe period) in the diffraction signal occurs over a reciprocal length $\Delta q=\frac{\pi}{R}$.

In our experimental geometry ($\lambda =\qty{1.24}{\nano\meter}$, pixel-size $\delta \theta = 0.0126\,^\circ$, this characteristic length can be expressed in pixels ($M$) for the average-sized particle used
\end{minipage}
\hfill
\begin{minipage}[t]{0.48\textwidth}
\setlength{\parindent}{1em}
\noindent in the main text and the maximum-sized particle detected:
\begin{align}
\vspace{-.2cm}
    M(\qty{354}{\nano\meter})&=8.0 ~\mathrm{ pixels}\\
    M(\qty{850}{\nano\meter})&=3.3 ~\mathrm{ pixels}
    \vspace{-.2cm}
\end{align}
A positional uncertainty of 1–2 pixels acts as a point spread function (PSF) that is significantly narrower than the minimum oscillation period of the signal. Thus, this localized shift preserves the structural information without significant blurring.

Furthermore, the sensitivity of real-space structural features to reciprocal-space uncertainties decreases at higher scattering angles. For a feature of size $x$ associated with a momentum transfer $q$, the relationship $x=\frac{2\pi}{q}$ implies, via error propagation, that the real-space uncertainty $\Delta x$ scales as:
\begin{align}
\vspace{-.2cm}
    \Delta x = \left| \frac{2 \pi}{q^2}\right| \Delta q
    \vspace{-.2cm}
\end{align}
For a fixed detector pixel uncertainty $\Delta q$, the resulting error in the estimated structural size $\Delta x$ is reduced by a factor of $q^2$. For example, a one-pixel error at a scattering angle of $0.3\,^\circ$ corresponds to an uncertainty of $\sim\qty{10}{\nano\meter}$, whereas the same error at $3\,^\circ$ (high-$q$) corresponds to only $\sim\qty{1}{\angstrom}$. Thus, the high-resolution region of the detector is inherently more robust to absolute pixel misplacements.

On top of these purely mathematical considerations, the $\qty{1}{\percent}$ bandwidth of the FEL introduces an intrinsic coordinate jitter in the scattering angle. At high scattering angles (e.g., $1.5\,^\circ$), this jitter is $\sim 1~\mathrm{ pixel}$, making the cluster-assignment shift commensurate with the existing physical limits of the experiment.
\end{minipage}

\vfill

\newlength{\bottomfigheightleft}
\newlength{\bottomfigheightright}

\setlength{\bottomfigheightleft}{0.21\textheight}
\setlength{\bottomfigheightright}{0.23\textheight}

\vspace*{\fill}

\noindent
\begin{minipage}[t]{0.48\textwidth}
    {\centering
    \includegraphics[height=\bottomfigheightleft]{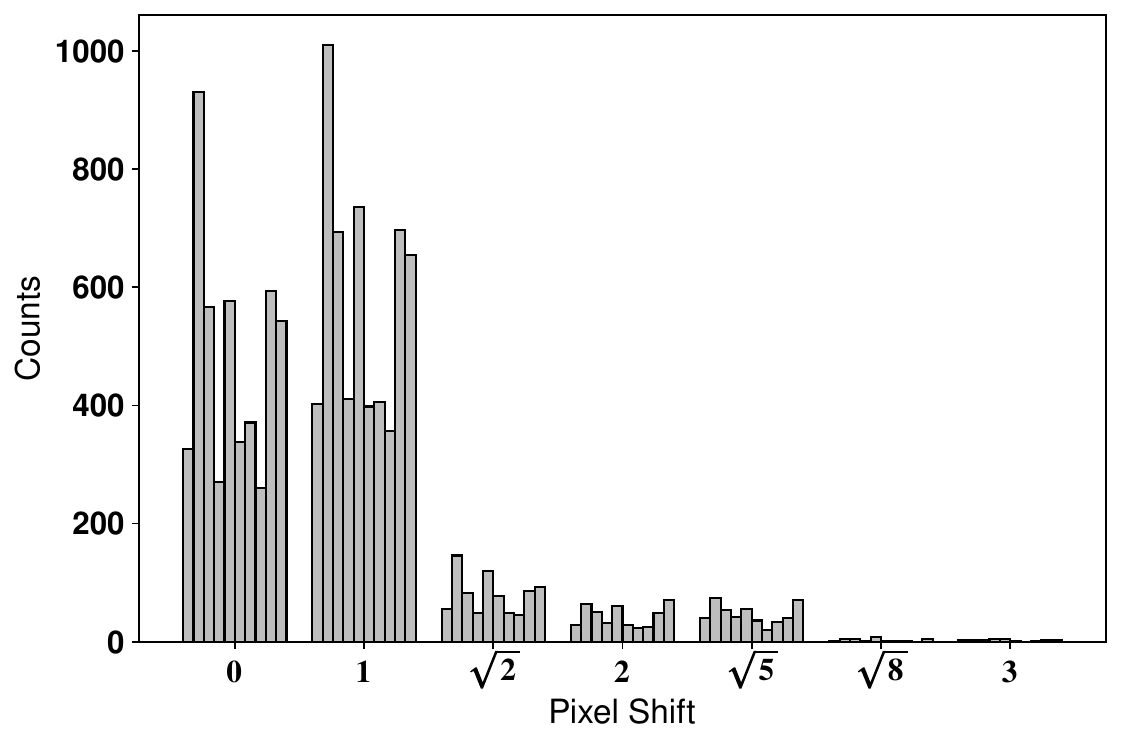}\par}

    \refstepcounter{figure}
    \label{fig:upper_bound_for_shift}

    \vspace{0.3em}
    {\small
    \noindent\textbf{Supplementary Fig.~\thefigure.}
    Upper bound on the distance between the edge of a photon cluster and its assigned pixel position. Ten randomly selected diffraction patterns were analyzed.
    \par}
\end{minipage}
\hfill
\begin{minipage}[t]{0.48\textwidth}
    {\centering
    \includegraphics[height=\bottomfigheightright]{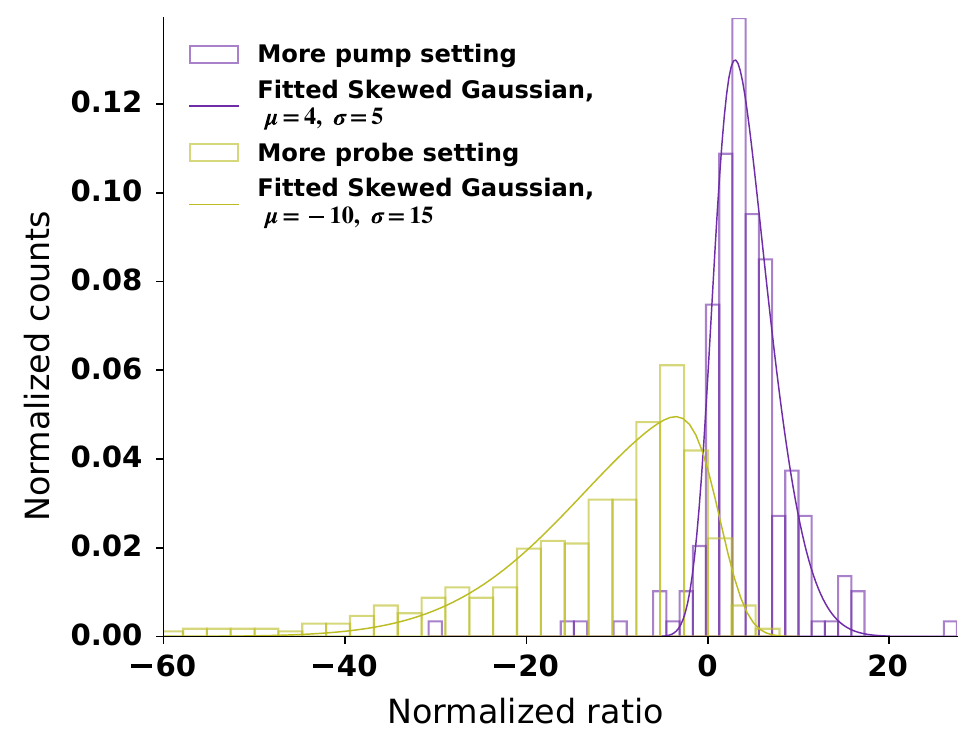}\par}

    \refstepcounter{figure}
    \label{fig:pulseRatDist}

    \vspace{0.3em}
    {\small
    \noindent\textbf{Supplementary Fig.~\thefigure.}
    Abundance distributions of the ratios of intensities of the two x-ray pulses, $I_\mathrm{red,\,blue}$ as determined from the pixel-based separation of patterns. The positive side of the $x$-axis shows the abundance of ratios $I_\mathrm{red}/I_\mathrm{blue}-1$ where $I_\mathrm{red}/I_\mathrm{blue}>1$, the negative side shows $1 - I_\mathrm{blue}/I_\mathrm{red}$ where $I_\mathrm{red}/I_\mathrm{blue}<1$. At $x=0$, $I_\mathrm{red}/I_\mathrm{blue}=1$. The data are fitted with a skewed Gaussian function, with $\mu$ describing the position of the distribution and $\sigma$ its characteristic width.
    \par}
\end{minipage}

\endgroup

\clearpage
\twocolumngrid